\documentclass[journal]{IEEEtran}

\usepackage{algorithmic}

\usepackage{cite}
\usepackage{graphicx}
\graphicspath{ {images/} }
\usepackage{todonotes}
\usepackage{amsmath}
\usepackage{listings}
\usepackage{subfigure}
\usepackage{multirow}

\usepackage[linesnumbered,ruled,vlined]{algorithm2e}

\begin{document}

\title{ Efficient Analog CAM Design}

\author{Jinane Bazzi, Jana Sweidan, Mohammed E. Fouda, Rouwaida Kanj, and Ahmed M. Eltawil
\thanks{J. Bazzi is with King Abdullah University of Science and Technology (KAUST), Thuwal 23955, Saudi Arabia.}
\thanks{J. Sweidan and R. Kanj are with ECE Dept., American University of Beirut, Lebanon, 1107 202}
\thanks{M. Fouda is with Center for Embedded \& Cyber-physical Systems, University of California-Irvine, Irvine, CA, USA 92697-2625. Email: foudam@uci.edu}
\thanks{A. Eltawil is with King Abdullah University of Science and Technology (KAUST), Thuwal 23955, Saudi Arabia.}
\thanks{The authors are thankful to King Abdullah University of Science and Technology (KAUST), Kingdom of Saudi Arabia, for funding this work.}
\thanks{Manuscript received xxx, xxx; revised xxxx, xxx.}}

\markboth{IEEE Transactions on Circuits and Systems: Regular Papers,~Vol.~xx, No.~xx, xxx~xxxx}%
{Bazzi \MakeLowercase{\textit{et al.}}:  Efficient Analog CAM Design}

\maketitle

\begin{abstract}
Content Addressable Memories (CAMs) are considered a key-enabler for in-memory computing (IMC). IMC shows order of magnitude improvement in energy efficiency and throughput compared to traditional computing techniques. Recently, analog CAMs (aCAMs) were proposed as a means to improve storage density and energy efficiency. In this work, we propose two new aCAM cells to improve data encoding and robustness as compared to existing aCAM cells. We propose a methodology to choose the margin and interval width for data encoding. In addition, we perform a comprehensive comparison against prior work in terms of the number of intervals, noise sensitivity, dynamic range, energy, latency, area, and probability of failure.
\end{abstract}


\IEEEpeerreviewmaketitle

\section{Introduction}

\IEEEPARstart{O}{n} account of the imminent barrier to Moore’s law in CMOS technology, also known as the power wall, revolutionary approaches for systems integration are needed. The quest for more advanced novel nanoelectronics with low-power functionality and dense reconfigurable electronics integration has brought up a promising element: the memristor. As a prospective candidate, the memristor has substantial advantages, such as inexpensive manufacturing, ultrahigh density,  non-volatility, low power, and most importantly CMOS compatibility, which allows the advancement of Moore’s Law beyond the present silicon roadmap horizons \cite{Kang2012}. Memristors are programmable and exhibit a controllable hysteresis during operation making them a viable candidate for a vast majority of applications \cite{1083337}. These include non-volatile processors, highly programmable and self-adaptable analog/digital electronics, in-memory compute, resistive nanocomputing architectures, and synaptic neuromorphic networks \cite{Kang2012}. 

More recently, associative  and  approximate  computing  using resistive  memory  based  Ternary  Content  Addressable  Memory (TCAM) is  becoming increasingly popular. Content Addressable  Memories (CAM)  are storage devices that  can be  searched  in a parallel manner to  match the input data and return the address of the matching content. While CAMs require an exact match to the input, Ternary CAMs stand out as an extension where partial searches are possible, as they allow  “don’t care”  conditions on parts of the searched data \cite{8053201}. The highly parallel in-memory operation of  the TCAM results in very high throughput compare  operations  at  low latency, leading to commercial success in different applications including network  routing \cite{1041061,253403}, real-time  network  traffic  monitoring \cite{7468531},  and  access  control  lists (ACLs) \cite{patent}.  
Such architecture styles, though immensely useful, struggle with high power consumption and low density when implemented in standard CMOS technology. Various memristor-based TCAM design options have been proposed to tackle these concerns, benefiting from their non-volatility and high packing density. Nevertheless, in most memristor-based  CAM  designs, the memristor encodes only binary states, and its highly tunable analog conductance is not utilized. 

Accordingly, the memristor-based analog CAM (aCAM) was proposed to leverage the tunable conductance of memristive devices to store a range of values in each cell \cite{main}. Cells can cover overlapping wide ranges or small discrete ranges. This allows to compare an analog or encoded multi-bit input to the stored range to decide on a match or a mismatch and thus enhances memory storage capability. Given the aCAM’s ability to store wide intervals of continuous levels, new search functionalities in the analog domain are possible \cite{main}.
This along with multi-bit capabilities offer improved memory densities along with reduced power \cite{main}, which in turn allows the use of aCAMs for more generic applications. These applications include associative computing \cite{batcher, tracy2016towards, guo2011, guo2013}  which benefit from these properties as it usually suffers from limited memory densities and high power consumption of traditional CAMs. {Furthermore, in aCAM, joint intervals may result in a larger number of possible non-mutually exclusive states. Multi-bit CAM cells (MCAM) can also be implemented using other switching devices such as NAND Flash \cite{9125991}. A recent study \cite{materials} compared different types of switching devices including memristors, phase change, magnetorsistors, and FeFET devices. Memristor devices stood out in many key properties such as the maximum number of distinguishable states, switching speed, endurance, and retention, thereby qualifying them as good candidates for aCAM applications.  
More recently, the authors in \cite{9372119, 9286826, 9106766, 9474025} introduced HfO$_2$ based Ferroelectric ﬁeld-effect transistors (FeFET) aCAM and MCAM cells where FeFETs were used due to improved compactness, multi-level programmability, and density. They demonstrated the ability to encode 2-3 bits, i.e., upto 8 levels, with some overlap between adjacent levels in the presence of variability \cite{9372119}.}
 
\begin{figure}[!t] 
\centering
\includegraphics[width=0.9\columnwidth]{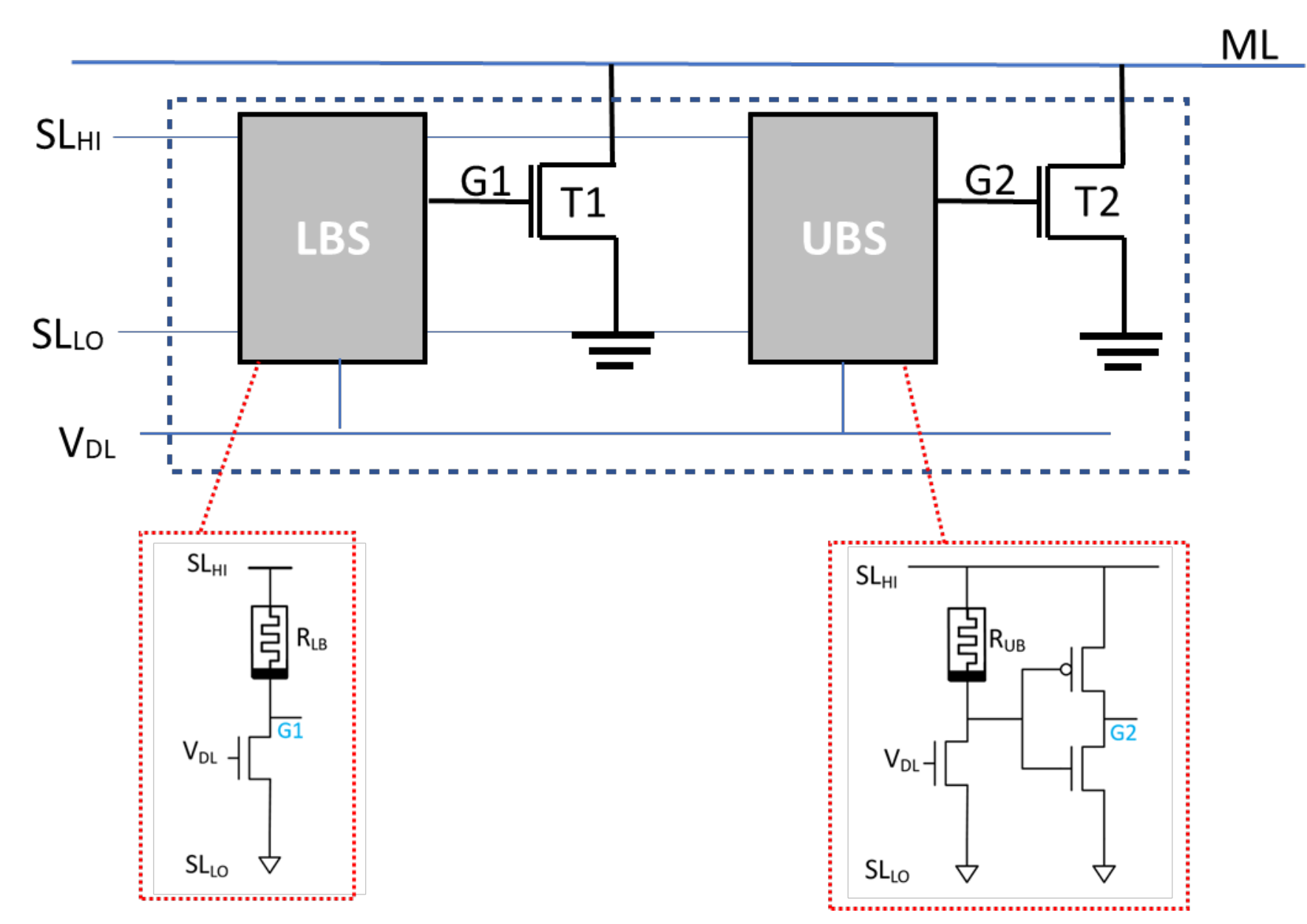}
\caption{Analog CAM cell structure showing the LBS and UBS of the 6T2M circuit.}
\label{fig:aCAM structure}
\vspace{-0.15in} 
\end{figure} 

\begin{figure} [!t]
\centering
\includegraphics[width=\columnwidth]{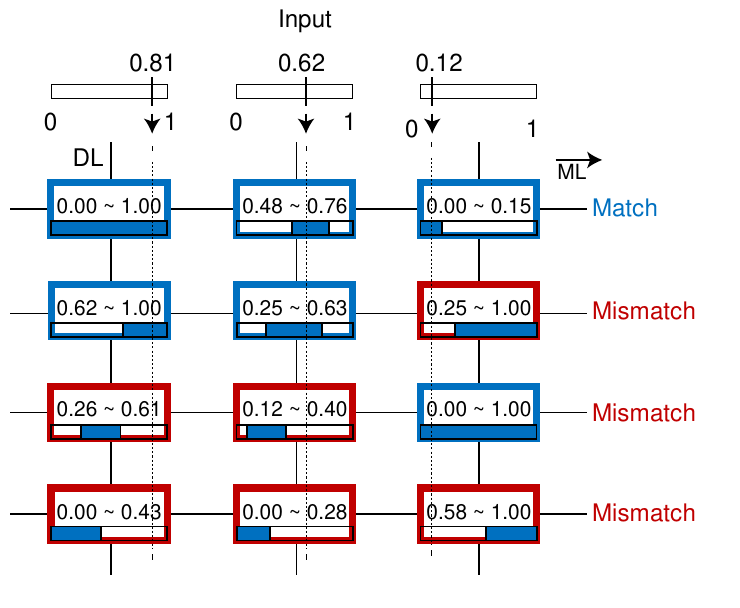}
\caption{Analog CAM search operation example. Blue colored cells represent the matched cells and red colored cells represent the mismatched cells, adopted from \cite{main}.}    
\label{fig:Search example}
\vspace{-0.2in} 
\end{figure}

{Our study is oriented towards the memristor based analog CAMs   analog CAMs with single data search lines and programmable analog/multi-bit intervals \cite{main}.} Fig. \ref{fig:aCAM structure} shows the structure of the memrsitor based aCAM cell proposed in \cite{main}. It consists of two voltage divider subcircuits: (1) the lower bound subcircuit (LBS) and (2) the upper bound subcircuit (UBS) that encode the stored interval lower and upper bounds, respectively. These stored bounds are configured by the programmed conductances of the memristors in each subcircuit as illustrated in Fig. \ref{fig:aCAM structure} for the 6-transistors 2-memristors analog CAM (6T2M) cell presented in \cite{main}. The search input data is mapped to voltage amplitudes $V_{DL}$ that is fed to the subcircuits.


The analog CAM \cite{main} is envisioned to enable an analog search operation where the input data to the aCAM is being searched against the stored intervals, as illustrated in Fig. \ref{fig:Search example}. The lower and upper bounds of these intervals represent the acceptance range for a match. Similar to digital CAMs, a full match only occurs when all the inputs to the row match the stored data in their respective cells in that row. Otherwise, we record a mismatch. Moreover, the cell can be used as digital CAMs if adjusted to store narrow ranges as discrete levels representing multi-bit operation. {For multi-bit applications, it is important to attain high gain for the cell designs to enable a plurality of discrete intervals with good differentiation between the match and mismatch scenarios. Furthermore, aCAM cells can be susceptible to process variations and noise along with other uncertainties \cite{9372119} which can be limiting to the underlying applications. In this paper, We propose two new aCAM cell designs with improved gain capabilities. We compare these designs along with a switch based cell \cite{main} in terms of different design metrics. We further develop an interval building algorithm for aCAM cells that draws boundaries, i.e., margins, between discrete intervals to ensure proper operation and guard against noise and process variations.}


The contributions of this paper can be summarized in the following:
\begin{itemize}
    \item {We propose two new aCAM cell designs: the 10T2M and 8T2M aCAM cells and explore the improvements in the gain and other design metrics in comparison to the 4T2M2S cell proposed in \cite{main}.}
    
    \item {For purposes of multi-bit storage, we develop an interval building algorithm to determine the memristor configurations that result in a maximum number of discrete intervals while maintaining conservative margins between the intervals to ensure proper differentiation between match and mismatch and ensure operation in the presence of noise and/or process variations. We relied on SPICE simulations for interval building and validation.} { We also relied on heuristics to determine for a desired number of intervals the configurations that maximize the dynamic margin, the match line difference between the full match and 1-mismatch scenarios.} 
    \item We compare the different designs in terms of key metrics such as the maximum number of intervals, dynamic range, latency, fail probability, energy, and area. Our studied memristor based CAMs proved the ability to encode up to 4 bits of storage and a maximum of 17 intervals {for the 8T2M cell and 24 intervals for 10T2M with 5-10x lower latency than the 4T2M2S design. The latter enabled 6 distinguishable intervals at the same noise guarding level (10\% $V_{DD}$) and upto 31 intervals at low noise guarding levels (2\% $V_{DD}$).}
    \item Finally, we study the behavior of the proposed aCAM cells under variability and process variation considerations. 
\end{itemize}

The remainder of the paper is organized as follows: Section II discusses the aCAM concept, operation, and the circuit realization that has been proposed in \cite{main}. Section III introduces two new circuit realizations for the aCAM and discusses the operation. Section IV examines the margin methodology and the algorithms implemented for interval building and figure of merit analysis. Section V presents the simulation results and a comparative analysis of the different circuit designs. Finally, section VI concludes the work.

\section{aCAM cell Design and Analysis}

\subsection{6T2M aCAM cell Overview}
\label{cell_overview}
Similar to current CAM setups, the search operation in the 6T2M design starts by precharging the ML to a high logic level, and the ML stays high (match) only when the search input is within the bounds determined by the two subcircuits, otherwise, ML discharges through transistors T1 and/or T2, thereby leading to a low logic level (mismatch) on the ML. Thus, a match occurs when both gate voltages of transistors T1 and T2, $V_{G1}$ and $V_{G2}$, respectively, are smaller than the threshold voltage, thereby keeping the transistor channels in a high resistance state. 

\textbf{Lower and Upper Bound Subcircuits:} as shown in Fig. \ref{fig:aCAM structure}, the LBS consists of a resistor-based inverter comprising a transistor and a series-connected memristor M1, which generates the gate voltage $V_{G1}$ for T1  to embody the aCAM cell’s lower bound LB match threshold. The value of this match threshold is conﬁgured by tuning the memristor conductance in the resistor-based inverter circuit. When an applied input voltage $V_{DL}$ is larger than this match threshold, a small voltage will build on G1 and yield a match due to the fact that T1 will remain off. The upper bound of the search range, UB, is conﬁgured similarly with an independent voltage divider using memristor M2 and an inverter to control the gate voltage $V_{G2}$ of the second pull-down transistor T2 (Fig. \ref{fig:aCAM structure}). As a result, the cell yields a match and keeps ML high only when $V_{DL}$ is within a certain range set by the resistance values for memristors $R_{LB}$ and $R_{UB}$. 



\begin{table} [!t]
\centering
\caption{Match-Mismatch Summary}
\renewcommand\arraystretch{1.5}
\begin{tabular}{|c|c|c|c|c|}
\hline
    \boldmath{$V_{G1}$}   &  \boldmath{$V_{G2}$}    & \textbf{T1}   & \textbf{T2}    & \textbf{Result}    \\  \hline
    $<V_{tn}$  &  $<V_{tn}$   & OFF  & OFF   & Match     \\  \hline
    $<V_{tn}$  &  $>V_{tn}$   & OFF  & ON    & UB Mismatch  \\  \hline
    $>V_{tn}$  &  $<V_{tn}$   & ON   & OFF   & LB Mismatch  \\  \hline
    
\end{tabular}
\label{tab:match/mismatch}
\end{table}

\begin{figure} [!t]
    \centering
       \subfigure[]{\includegraphics[width=0.75\columnwidth]{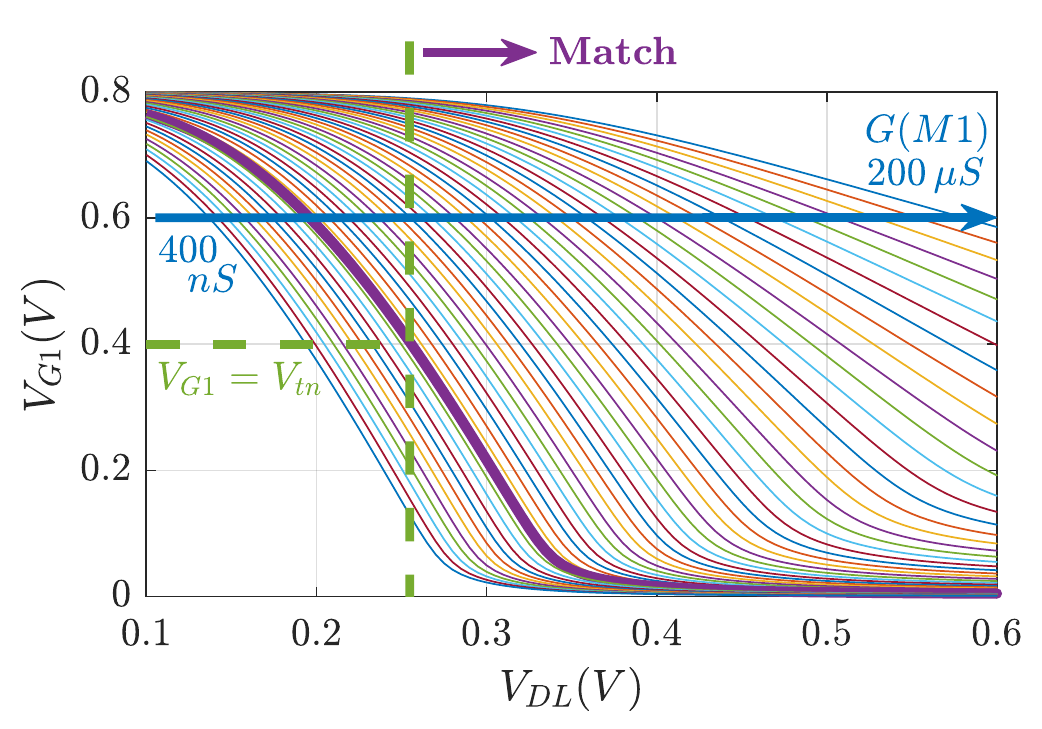}}
       \subfigure[]{\includegraphics[width=0.75\columnwidth]{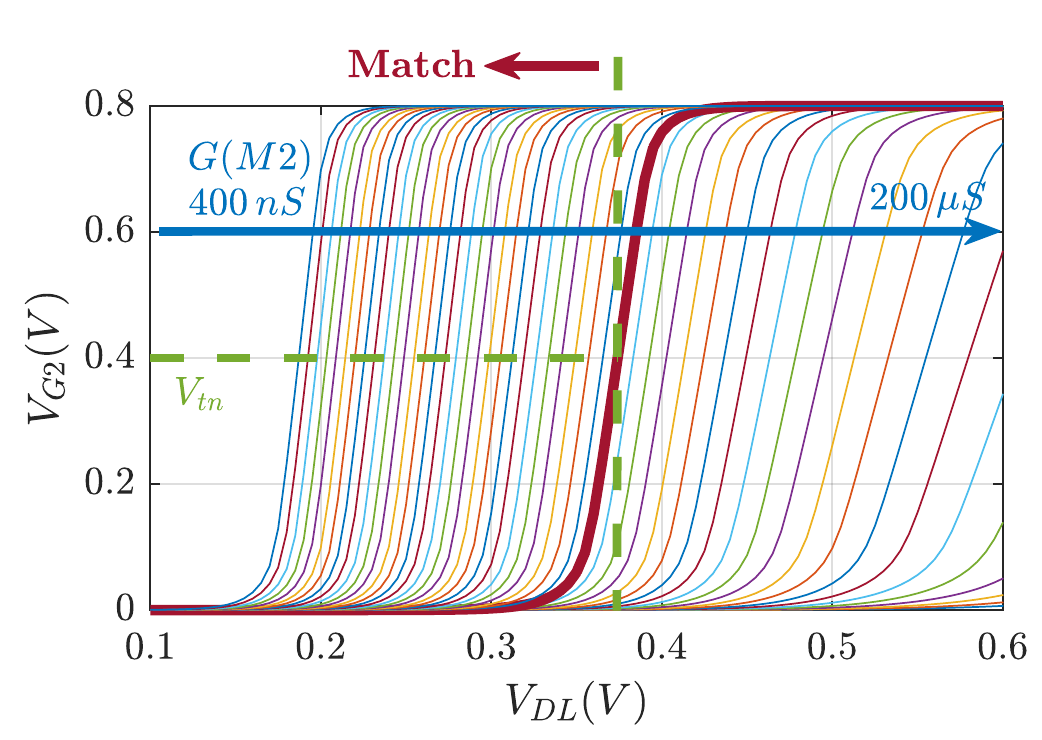}}
    \caption{ Voltage transfer characteristics (a) $V_{G1}$ vs $V_{DL}$ and (b) $V_{G2}$ vs $V_{DL}$ for $R \in [5K\Omega, \, 2.5M\Omega]$.}
    \label{fig:VG vs VDL}
\vspace{-0.15in} 
\end{figure}

Table \ref{tab:match/mismatch} summarizes the match/mismatch states based on $V_{G1}$ and $V_{G2}$ values in an ideal framework. Fig. \ref{fig:VG vs VDL} represents the LBS and UBS voltage transfer characteristics (VTC) for $V_{G}$ versus $V_{DL}$. As such, we can rely on such curves to determine the lower and upper bounds for a given pair of M1 and M2 resistances, respectively, based on the $V_{DL}$ values that result in $V_{G1}=V_{tn}$ and $V_{G2}=V_{tn}$. For example, for the simulated VTC curves in Fig. \ref{fig:VG vs VDL}, the range: [LB, UB] = [0.255, 0.374]V can be obtained for $R_{LB}=619K\Omega$ and $R_{UB}=63.1K\Omega$. 

\subsection{Cell Conductance Sensitivity}
\label{cond sensitivity}
In \cite{main}, it was stated that when the word length is long enough, sensing errors and/or changes in the accepted search range of an aCAM cell are likely to occur. This is because the cell conductance, $G$, is related to current leakages through the pull-down transistors, whose gates are controlled by $V_{DL}$. So if $G$ is continuous with respect to $V_{DL}$ ($G = f(V_{DL}$)), the higher $\frac{\partial G_T}{\partial V_{DL}}$ at the boundary of interval, the better the capability to store more accurate ranges and accordingly more bits of discrete levels. $\frac{\partial G_T}{\partial V_{DL}}$ can be best defined as:

\begin{equation}
\frac{\partial G_T}{\partial V_{DL}} = (\alpha  S_s)^{-1}
\label{sensitivity}
\end{equation}
where $\alpha = \frac{\partial V_{DL}}{\partial V_G}$ = (1/gain), and $S_s$  is the subthreshold swing slope of the transistor. So for better sensitivity, we need lower $S_s$ and higher gain, i.e., steeper $V_G$ vs $V_{DL}$ VTC slopes, as shown in Fig. \ref{fig:VG vs VDL}. 

\subsection{4T2M2S aCAM Cell}
Accordingly, to enhance the sensitivity and improve the word length in the aCAM, the authors in \cite{main} proposed to rely on a volatile threshold switching memristors TS in place of transistors T1 and T2. These devices are characterized by sharp transitions between states  and small $S_s$ = 1mV/Dec \cite{ref3}. This enhances the conductance sensitivity (Eq. (\ref{sensitivity})). A smaller change in $V_{DL}$ will lead to a faster transition from $R_{ON}$ to $R_{OFF}$, hence greatly limiting leakages and improving column interference issues usually exposed in the 6T2M circuit design.

\begin{figure} [!t]
\centering
\includegraphics[width=0.9\columnwidth]{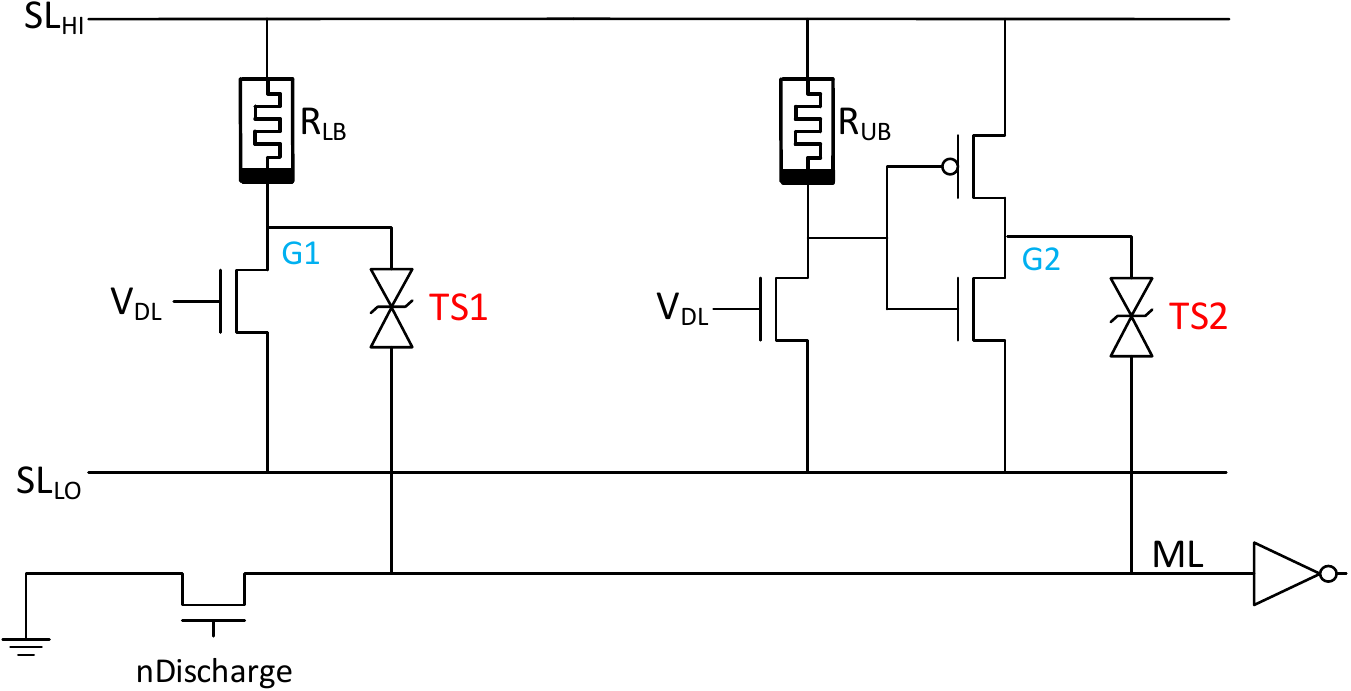}
\caption{Schematic of the 4T2M2S aCAM cell composed of four transistors, two meristors, and two threshold switch devices \cite{main}-Supplementary.}
\label{fig:aCAM TS}
\vspace{-0.15in} 
\end{figure}

The modified circuit that utilizes the threshold switching memristors is shown in Fig. \ref{fig:aCAM TS}. This design encompasses a discharge phase followed by an evaluate phase. Thus, in contrast to the 6T2M where the match line ML, initially charged, is discharged using T1/T2 for mismatch cases, a search operation for this design starts with ML at ground, and the ML is charged up using TS1/TS2 only for mismatch cases. When the cell is in matching state, the voltages $V_{G1}$ and $V_{G2}$ will be less than the switching threshold voltage of the TS device, $V_{T1}$, so TS1 and TS2 remain off, and ML remains 0. For the case of lower (upper) mismatch, $V_{G1}$ ($V_{G2}$) will be greater than $V_{T1}$, so TS1 (TS2) will turn on, which results in a path from $SL_{HI}$ through TS1 (TS2) to ML, thereby charging ML.

\section{Proposed aCAM DESIGN ALTERNATIVES}
In this section, we present new aCAM cell designs that improve the conductance sensitivity in (Eq. (\ref{sensitivity})) and hence the word length and memory storing capabilities through improving the gain as opposed to the 4T2M2S cell that relies on enhancing subthreshold swing. We focus on improving $\frac{\partial V_{G1}}{\partial V_{DL}}$ for the LBS design which demonstrates lower gains compared to $\frac{\partial V_{G2}}{\partial V_{DL}}$ in the UBS design as can be inferred from the VTC curves in Fig. \ref{fig:VG vs VDL} where the UBS VTC shows steeper transitions. To enhance the sensitivity of the LB curves and obtain steeper lower bounds (better sensitivity), we propose two new LBS designs. Note that the UBS remains unchanged.

\subsection{10T2M aCAM Cell}
\begin{figure} [!t]
    \centering
       \subfigure[]{\includegraphics[width=0.49\columnwidth]{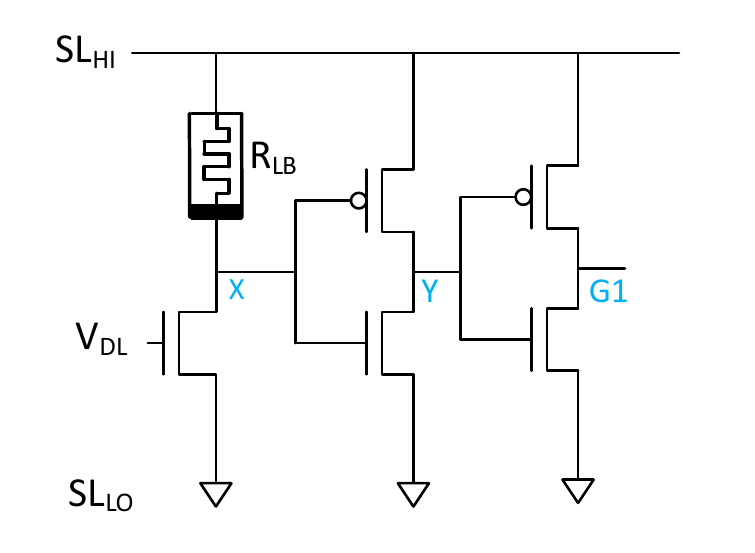}}
       \subfigure[]{\includegraphics[width=0.49\columnwidth]{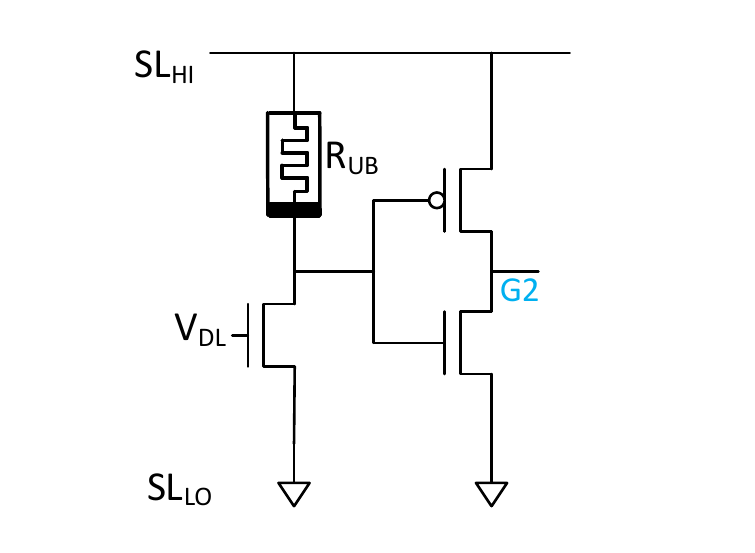}}
    \caption{LBS and UBS for the 10T2M aCAM (a) Lower bound subcircuit LBS and (b) Upper bound subcircuit UBS remains unchanged.}
    \label{fig:10T2M subcircuits}

\vspace{-0.15in} 
\end{figure}

\begin{figure} [!t]
    \centering
       \subfigure[]{\includegraphics[width=0.49\columnwidth]{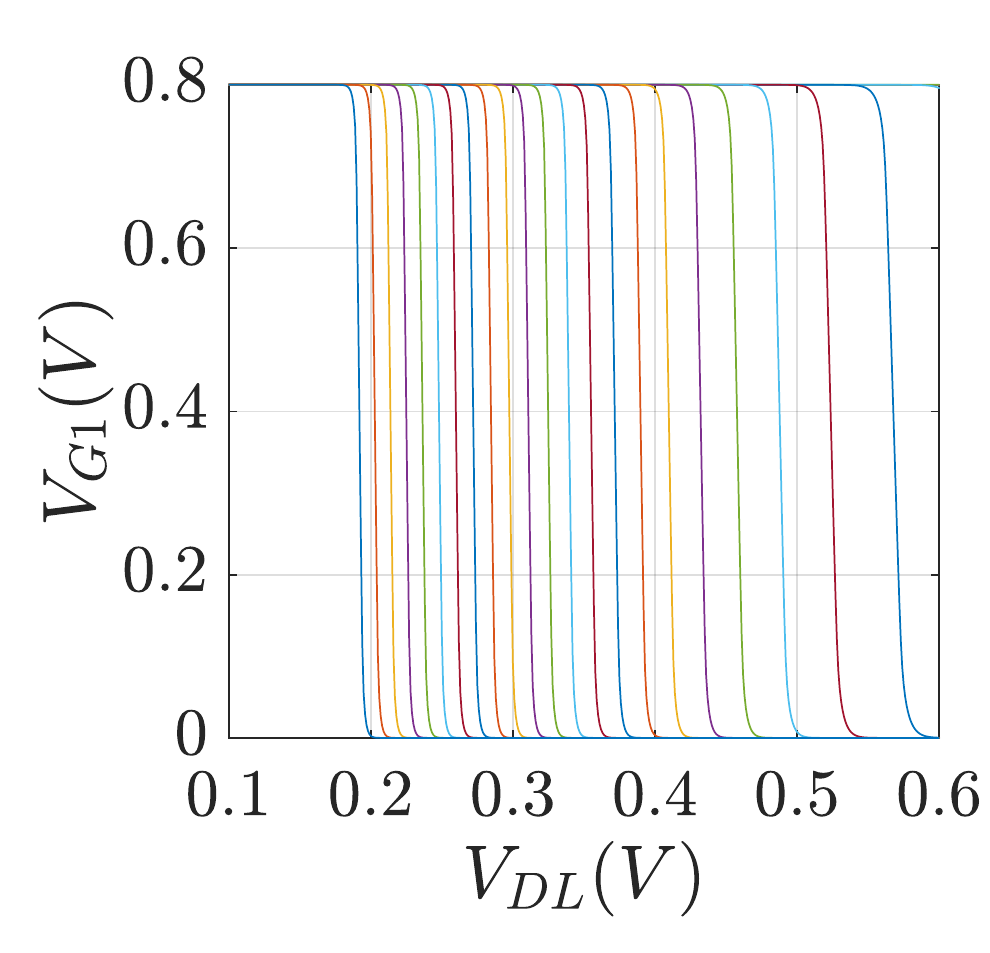}}
       \subfigure[]{\includegraphics[width=0.49\columnwidth]{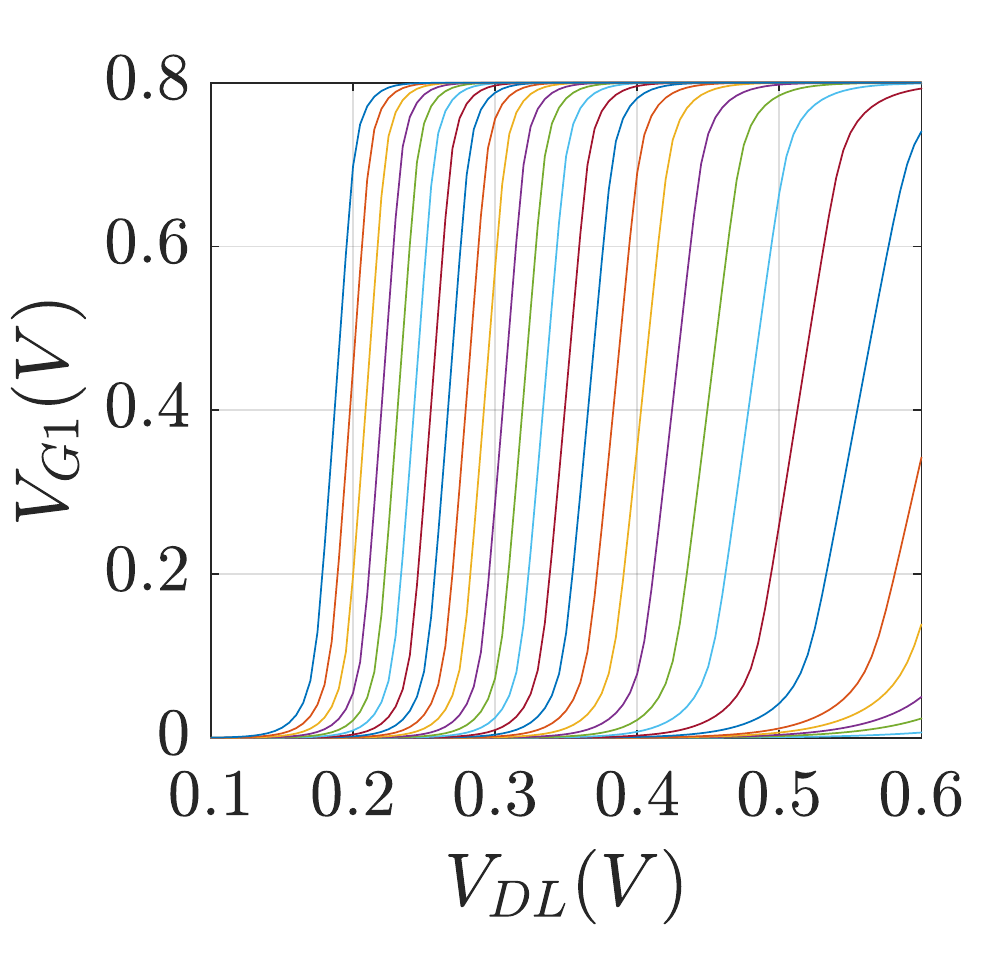}}
    \caption{Voltage transfer characteristics, $V_{G1}$ vs $V_{DL}$, of the lower bound circuit for (a) the 10T2M cell and (b) the 8T2M cell.}
    \label{fig:new designs VG1}

\vspace{-0.15in}
\end{figure}

\begin{figure} [!b]
\centering
\includegraphics[width=0.8\columnwidth]{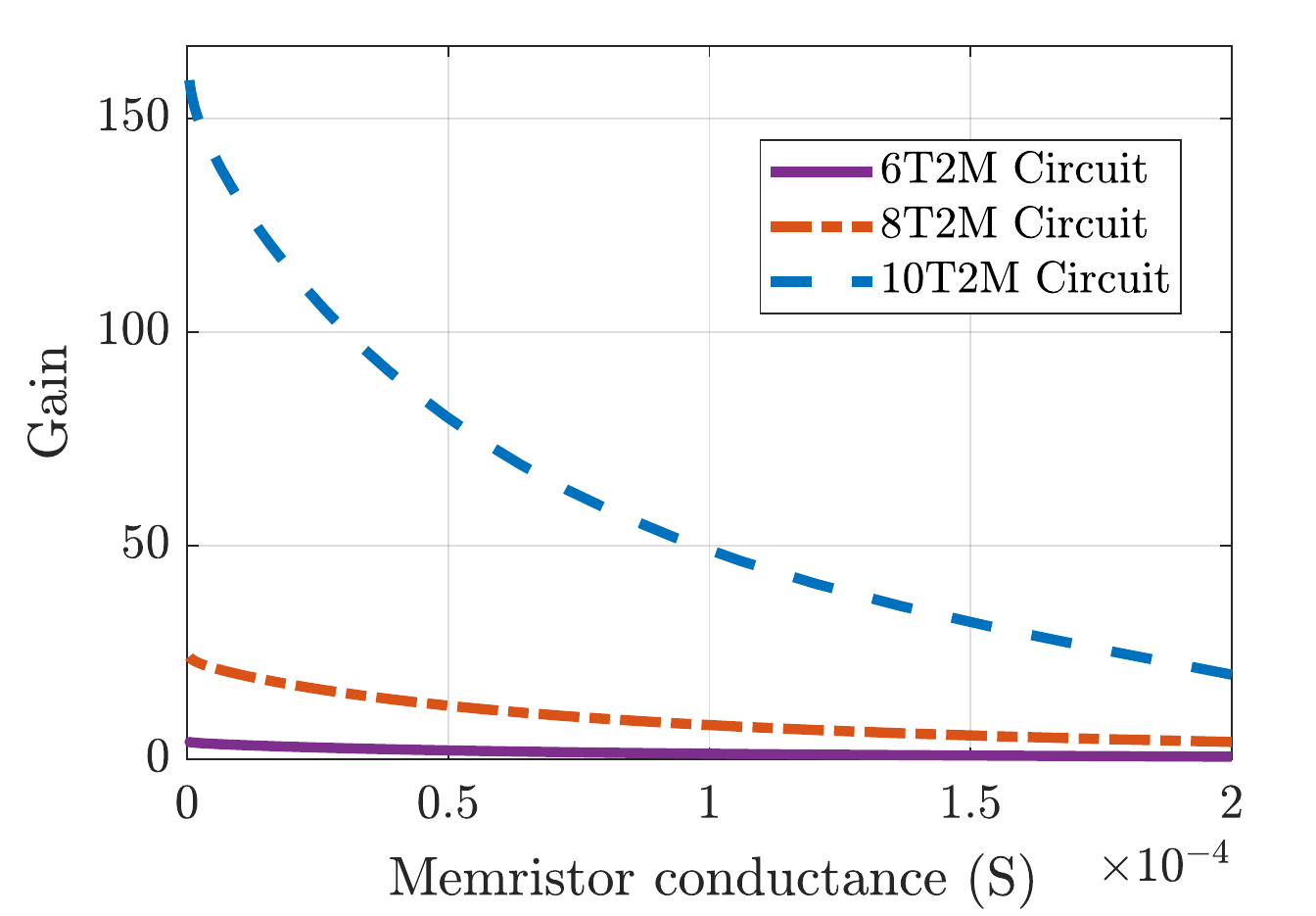}
\caption{Plot of gain vs memristor's conductance for the 10T2M, 8T2M, and 6T2M circuits.}
\label{fig:gain vs cond}
\vspace{-0.15in} 
\end{figure} 

We first study inserting a simple non-inverting buffer composed of two inverters before T1 to maintain the same functionality, while boosting gain (albeit at a cost of area) as illustrated in Fig. \ref{fig:10T2M subcircuits}. Fig. \ref{fig:new designs VG1} (a) illustrates the enhancement in the slopes of the LBS, as compared to that of the initial 6T2M circuit (Fig. \ref{fig:VG vs VDL} (a)), as they became much steeper. This results in up to 40 times improvement in the gain (at low conductance) compared to the 6T2M design (Fig. \ref{fig:gain vs cond}).

\subsection{8T2M aCAM cell}
\begin{figure} [!b]
\centering
\includegraphics[width=0.9\columnwidth]{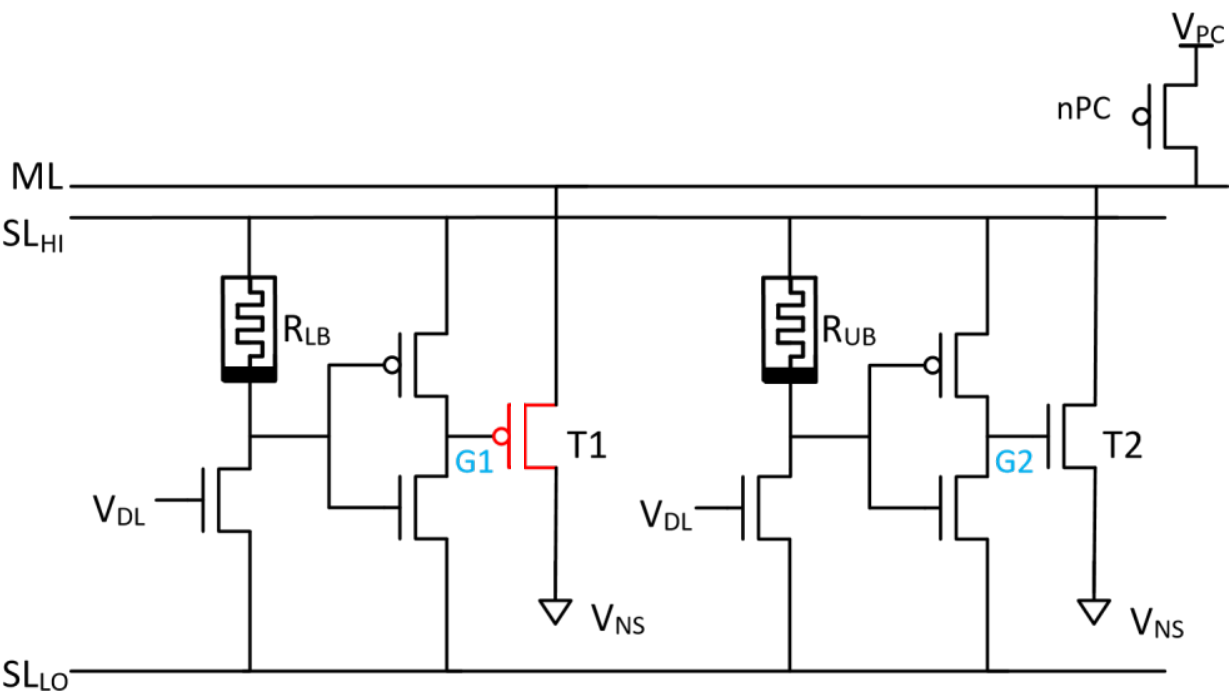}
\caption{Schematic of a complete 8T2M aCAM cell.}
\label{fig:aCAM 8T2M}
\vspace{-0.15in}
\end{figure}

A more compact version can be introduced by replacing T1 with a PMOS transistor preceded by an inverter as illustrated in Fig. \ref{fig:aCAM 8T2M}. We studied the VTC curves ($V_{G1}$ vs $V_{DL}$) as well as the gain in Fig. \ref{fig:new designs VG1} (b) and \ref{fig:gain vs cond}, respectively, for the 8T2M cell. We record up to 6 times improvement in the gain for this new design in comparison to the 6T2M cell.
Note that for the 8T2M, for a LB match, T1 (PMOS) must be off; hence, $V_{G1}$ must be high, $V_{G1}>V_{PC} - |V_{tp|}$, where $V_{PC}$ is the ML pre-charge voltage.  
Table \ref{tab:pmos} summarizes the match/mismatch states criteria for the 8T2M cell in an ideal framework. In addition, Some examples of transient signals of reading the three cells are shown in detail in the Supplementary materials (Fig.3).

\begin{table} [!b]
\centering
\caption{8T2M design LBS Match Mismatch Summary. Note that low $V_{G1}$ results in a lower bound mismatch.}
\begin{tabular}{|c|c|c|c|c|}
\hline
    \boldmath{$V_{G1}$}  &  \boldmath{$V_{G2}$}  &  \textbf{T1}  &  \textbf{T2}  &  \textbf{Result}   \\ \hline
    $<V_{PC}-|V_{tp}|$  &  $<V_{tn}$  &  ON  & OFF & LB Mismatch \\ \hline
    $>V_{PC}-|V_{tp}|$  &  $<V_{tn}$  &  OFF & OFF & Match    \\ \hline
    $>V_{PC}-|V_{tp}|$  &  $>V_{tn}$  &  OFF & ON  & UB Mismatch \\ \hline
\end{tabular}
\label{tab:pmos}
\end{table}

\section{INTERVAL BUILDING AND MARGINING}

{ Herein, we develop an interval building algorithm that determines the proper memristor configurations in order to maximize the number of aCAM discrete intervals without sacrificing functionality while guarding against noise and process variations.}

\subsection{aCAM Row}
Each aCAM cell creates a pull-down (up) path to the ML based on its state: match or mismatch. Thus, the cell is associated with a corresponding effective resistance, and hence channel conductances of transistors (or switching memristors) involved. 

\begin{figure} [!t]
    \centering
        \includegraphics[width=0.9\columnwidth]{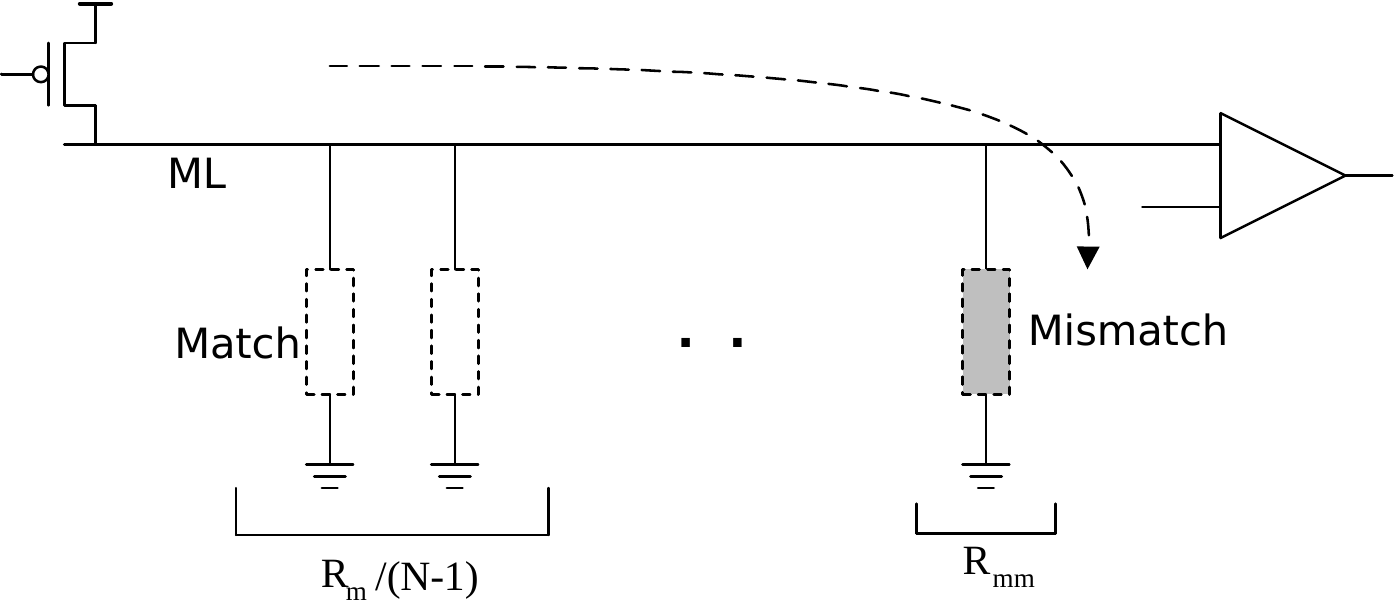}
    \caption{Schematic of an analog CAM word consisting of N cells with only one-bit in mismatch during the search operation. $R_{OFF}$ and $R_{ON}$ are the effective resistors of a match cell and a mismatch cell, respectively.}
    \label{fig:aCAM row}
\vspace{-0.15in} 
\end{figure}

The effective match and mismatch resistances of an aCAM cell are $R_{m}=0.5R_{off}$ and $R_{mm}=R_{on}//R_{off}\sim R_{on}$, where $R_{on}$ and $R_{off}$ are the on and off resistances of the pull-down transistor for the 10T2M and 8T2M cells, and the effective resistances for the 4T2M2S cell. As several cells are connected on the same ML in a row, just like digital CAMs, a row ML outputs “high” (“low” for 4T2M2S circuit) when all the cells in the row match. This represents a full match state (fm). A mismatch state incorporates charging (discharging for 4T2M2S) the ML. We typically optimize the design to properly differentiate between the fm and the one mismatch state (1mm) being the weakest mismatch case as shown in Fig. \ref{fig:aCAM row}.

The search range for one cell is affected by other cells in a row. Subthreshold current leakages via the ML pull-down transistors usually restrict the maximum length of CAM words and the number of bits held by each cell. Given $N$ cells per row, the fm and 1mm effective resistances, $R_{fm}$ and $R_{1mm}$, respectively, of the aCAM row can be defined as follows:

\begin{subequations}
\begin{equation}
R_{fm} = \frac{R_{m}}{N}
\end{equation}
\begin{equation}
R_{1mm} =\frac{R_m R_{mm}}{R_m +(N-1) R_{mm}}
\end{equation}
\end{subequations}

The design is robust when operating in conditions that enable differentiation between $R_{1mm}$ and $R_{fm}$. This can be either achieved by choosing sparse discrete levels, or by relying on designs whose effective resistance or conductance demonstrate high sensitivity to changes in $V_{DL}$. This is in coherence with the ($\frac{\partial G_T}{\partial V_{DL}}$) requirement in section \ref{cond sensitivity}.

\subsection{Margining Overview}

\begin{table} [!b]
\centering
\caption{Basic definitions for discrete level programming}
\begin{tabular}{|c|l|}
\hline
    \boldmath{$\eta$}  & \begin{tabular}[c]{@{}l@{}}number of discrete voltage levels; e.g., $\eta=8$ represents 3-bit \\ storage in the cell   \end{tabular} \\ \hline
    
    \boldmath{$D_i$}  & represents the $i^{th}$ discrete voltage level. $D_i\in[0,V_{DD}]$   \\ \hline
    
    \boldmath{$R_{{LB}_i}$}  & \begin{tabular}[c]{@{}l@{}} represents LB subcircuit memristor value corresponding to \\ the $i^{th}$ discrete voltage level $D_i$  \end{tabular} \\ \hline
   
    \boldmath{$R_{{UB}_i}$}  & \begin{tabular}[c]{@{}l@{}}  represents UB subcircuit memristor value corresponding to \\ the $i^{th}$ discrete voltage level $D_i$  \end{tabular}  \\ \hline
    
    \boldmath{$Cell_i$}  &  cell that is programmed using $R_{{LB}_i}$ and $R_{{UB}_i}$   \\ \hline
\end{tabular}
\label{tab:definitions1}
\end{table}

As stated in \cite{main}, the memristor configurations can be used to allow the array to store “continuous ranges of value”, or “discrete levels”. In the context of our work, we are interested in identifying memristor configurations for discrete levels. Based on this, we assume that the array cells may be storing discrete values for purposes of multi-bit storage. We determine basic definitions for discrete level programming given in Table \ref{tab:definitions1}.

Our objective is to identify functional levels $\{D_i\}$, and their corresponding programmable values, {$R_{{LB}_i}$} and {$R_{{UB}_i}$}, that maximize the potential of a given design without sacrificing accuracy and functionality. Particularly, the intervals must allow for proper differentiation between full-match and one mismatch states, i.e., differentiating between $R_{fm}$ and $R_{1mm}$. This can be achieved when $R_{m} \gg R_{mm}$, and the ideal set $\{D_i\}$ allows for large $R_{mm}$ during match for minimal leakages, and small $R_{m}$ of the mismatching cells for fast evaluations during mismatch and proper separation. 

\begin{figure} [!t]
\vspace{-0.1in} 
\centering
\includegraphics[width=0.85\columnwidth]{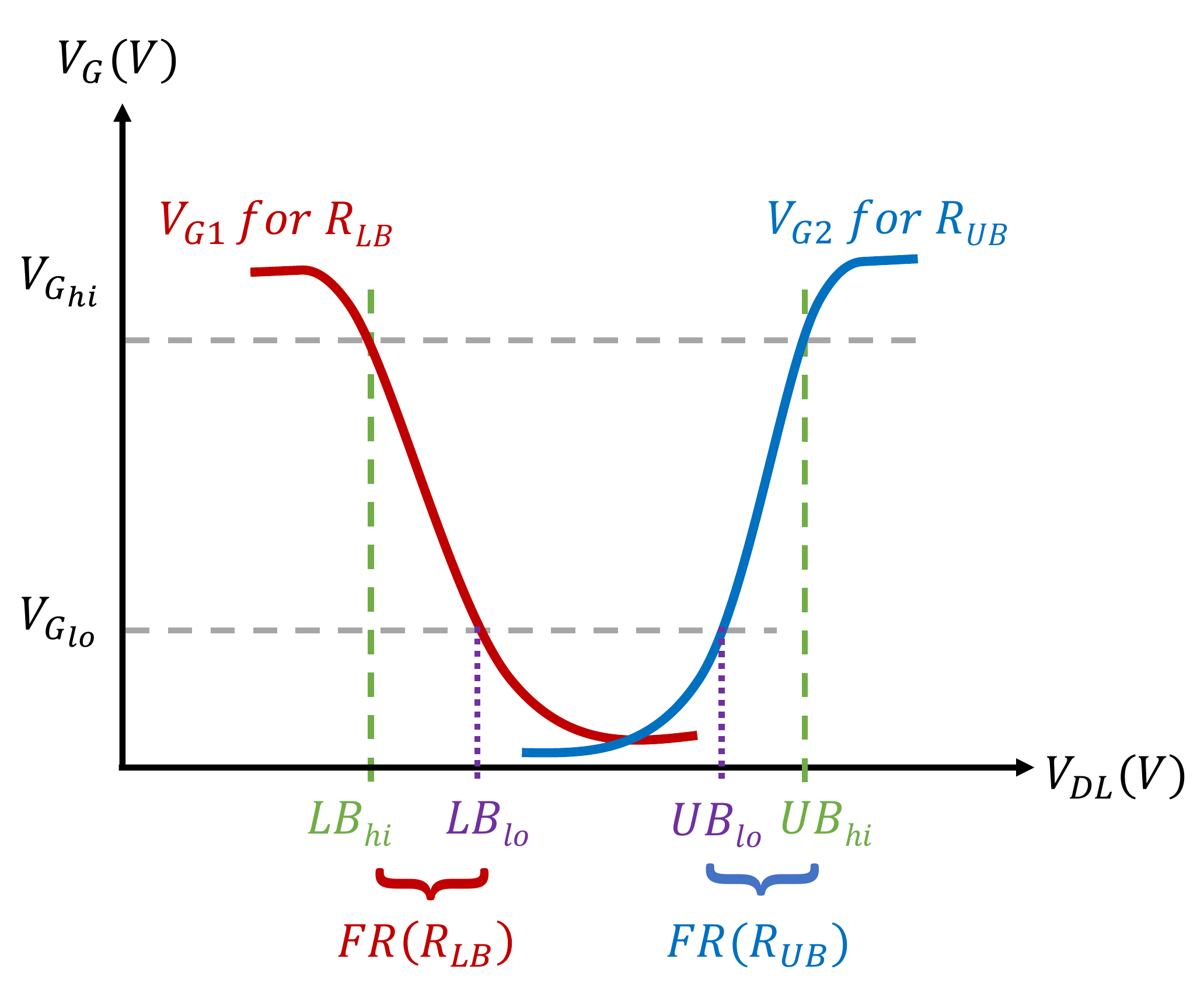}
\caption{LBS and UBS VTC for an arbitrary cell.}
\label{fig:VG1 and VG2}
\end{figure}

{In the simplest form, to determine the stored intervals, we can assume a cutoff voltage equal to the threshold voltage of the transistors (e.g., $V_{tn}=0.4=V_{DD}/2$). Thus, for a given pair of $R_{LB}$ and $R_{UB}$ resistances, the LB and UB of the stored interval are defined as the $V_{DL}$ values that result in $V_{G1}=V_{DD}/2$ and $V_{G2}=V_{DD}/2$, respectively, as explained in section \ref{cell_overview}. However, this eliminates the differentiation between the effective $R_m$ and $R_{mm}$. 
Furthermore, the designs can be subject to noise and/or process variations and this can affect their operation.}
\\{To guarantee proper functionality in terms of the ability to differentiate between match and mismatch scenarios and to guard against noise and process variations, we devise the following. Instead of relying on a single cutoff level to build the intervals, we rely on two cutoff levels, as will be explained next, that result in margins between the intervals.}
{These margins help us restrict to specific levels the outputs of the LB and UB subcircuits, which in turn are inputs to the devices that discharge the match line and hence help maintain reasonable effective $R_m$ and $R_{mm}$.}
{\textit{This guarantees (1) better dynamic ranges and better responses.} As such, uncertainties are minimized and a matching value for one interval results in a proper mismatch for another. It also \textit{(2) guards against noise and process variations.} For the latter, we  incorporate Monte Carlo analysis for forbidden region construction to secure proper functionality for non-nominal cells with process variations.}
\newline
{\textbf{Definitions:} To build the intervals with proper margins, we rely on the following terms that are key for our interval building algorithm} { as summarized in Table \ref{tab:definitions2}, and illustrated in Fig. \ref{fig:VG1 and VG2}. The figure presents a sketch for the lower bound and upper bound subcircuit output responses $V_{G1}$-$V_{DL}$ and $V_{G2}$-$V_{DL}$ respectively for an arbitrary $R_{LB}$ and $R_{UB}$; output nodes $G_1$ and $G_2$ feed the switches that will charge or discharge the match line.}

{The list of terms starts with the two subcircuit output cutoff levels that will be used for the interval building: $V_{G_{lo}}$ and $V_{G_{hi}}$ as presented next along with some other definitions.}

\begin{enumerate}

    \item {$V_{G_{lo}}$ represents the desired subcircuit output response low voltage levels as a percentage of $V_{DD}$ for proper match. E.g., $V_{G_{lo}}=0.1*V_{DD}$.}
    
    \item {$V_{G_{hi}}$ represents the desired subcircuit output response high voltage levels as a percentage of $V_{DD}$ for proper mismatch. E.g., $V_{G_{hi}}=0.9*V_{DD}$.}
    
    \item {$LB_{lo}$: corresponds to $V_{DL}$ value that results in $V_{G1}=V_{G_{lo}}$ for the LB circuit.}
    
    \item {$LB_{hi}$: corresponds to $V_{DL}$ value that results in $V_{G1}=V_{G_{hi}}$ for the LB circuit.}
    
    \item {$UB_{lo}$: corresponds to $V_{DL}$ value that results in $V_{G2}=V_{G_{lo}}$ for the UB circuit.}
    
    \item {$UB_{hi}$: corresponds to $V_{DL}$ value that results in $V_{G2}=V_{G_{hi}}$ for the UB circuit.}
    
    \item {$[LB_{lo}, UB_{lo}]$: Match interval boundaries guarantee $V_{G1}$ and $V_{G2}$ values to be low enough for interval match. This ensures better matches and leaves room or noise guarding.}
    
    \item {$FR(R_{LB})=[LB_{hi},\, LB_{lo}]$: Lower forbidden region boundary  guarantees $V_{G1}$ output to be high enough for inputs from lower intervals. This ensures better mismatches and leaves room for noise.}
    
    \item {$FR(R_{UB}) =[UB_{lo},\, UB_{hi}]$: Upper forbidden region boundary guarantees $V_{G2}$ output to be high enough for inputs from higher intervals. This ensures better mismatches.}
\end{enumerate}


\textbf{Interval match mismatch requirements:} {The corresponding matching levels $\{V_{DL}\}$ must be restricted within values $[LB, \, UB] = [LB_{lo}, \, UB_{lo}]$. This results in low enough $V_{G1}$ and $V_{G2}$ such that the switches are off and the defined matching interval is associated with large $R_{m}$ for a given matched cell.} As such, we define the discrete level, $D$, associated with a given pair ($R_{LB}$, $R_{UB}$) for a given cell to be according to: 

\begin{equation}
D = \frac{LB_{lo}+UB_{lo}}{2}
\label{Dval}
\end{equation}

We also have forbidden regions, as defined in items (8) and (9) above, surrounding a given predefined interval [$LB_i$, $UB_i$]. No other interval can start within those forbidden regions. Hence, with this requirement for given input ${D_j}, \, (j \neq i)$, the cell is guaranteed to operate in strong mismatch state. As such, the forbidden regions for a given cell encompass the $V_{DL}$ values that result in $V_{G_{lo}} < V_{G1,2} < V_{G_{hi}}$. $V_{DL}$ beyond the forbidden regions will result in properly turning on the switches of the cell. 
\begin{table}
\centering
\caption{Interval Definitions}
\resizebox{\columnwidth}{!}{%
\begin{tabular}{|c|l|}
\hline
\boldmath{$V_{G_{lo}}$}  &  
Target output low level  \\ \hline
\boldmath{$V_{G_{hi}}$}  &  

Target output high level \\ \hline
     \boldmath{$LB_{lo}(R_{LB})$}  & $V_{DL}\ |\ V_{G1}(R_{LB})=V_{G_{lo}}$ \\ \hline
    
     \boldmath{$UB_{lo}(R_{UB})$}  &  $V_{DL}\ |\ V_{G2}(R_{UB})=V_{G_{lo}}$  \\ \hline
    
     \boldmath{$LB_{hi}(R_{LB})$}  &$V_{DL}\ |\ V_{G1}(R_{LB})=V_{G_{hi}}$  \\ \hline
     
     \boldmath{$UB_{hi}(R_{UB})$}  & $V_{DL}\ |\ V_{G2}(R_{UB})=V_{G_{hi}}$  \\ \hline
     \boldmath{$FR(R_{LB}) $}  &  \begin{tabular}[c]{@{}l@{}} $[LB_{hi}(R_{LB}), LB_{lo}(R_{LB})]$
     \\ Left region: ${V_{DL}\ |\ V_{G_{lo}}<V_{G1}(R_{LB})<V_{G_{hi}}}$\end{tabular} 
\\ \hline
     \boldmath{$FR(R_{UB})$}  &  \begin{tabular}[c]{@{}l@{}} $[UB_{hi}(R_{UB}), UB_{lo}(R_{UB})]$
    \\ Right region: ${V_{DL}\ |\ V_{G_{lo}}<V_{G2}(R_{UB})<V_{G_{hi}}}$ \end{tabular} 
\\ \hline
\end{tabular}
}
\label{tab:definitions2}
\end{table}

\subsection{Interval Defining Algorithm}
Depending on the requirements set for $V_{G_{lo}}$ and $V_{G_{hi}}$, the number of feasible intervals within a given range can vary. Relaxed requirements result in increased number of intervals, however, this can be limiting in terms of the array size. In fact, the authors in \cite{main} demonstrate that the aCAM cell can support up to 20 discrete levels for an array of 512x64 columns. They also state that the number of discrete levels will vary based on the size of the array and particularly the number of columns in the arrays.

In what follows, we present an interval defining algorithm as function of $V_{G_{lo}}$ and $V_{G_{hi}}$ {that, given a range of memristor values, determines for a given design the maximum number of discrete intervals of width=W that can be stored in aCAM cells along with their corresponding memristor configurations while maintaining a predefined conservative margin between the intervals to ensure proper operation}. This will be used to study and compare the different designs subject to a set of critical performance metrics in terms of latency, leakage, and encode capability as function of the array size. The algorithm can be best defined as follows. It is split into two phases;  

\begin{figure} [!t]
    \centering
        \includegraphics[width=0.85\columnwidth]{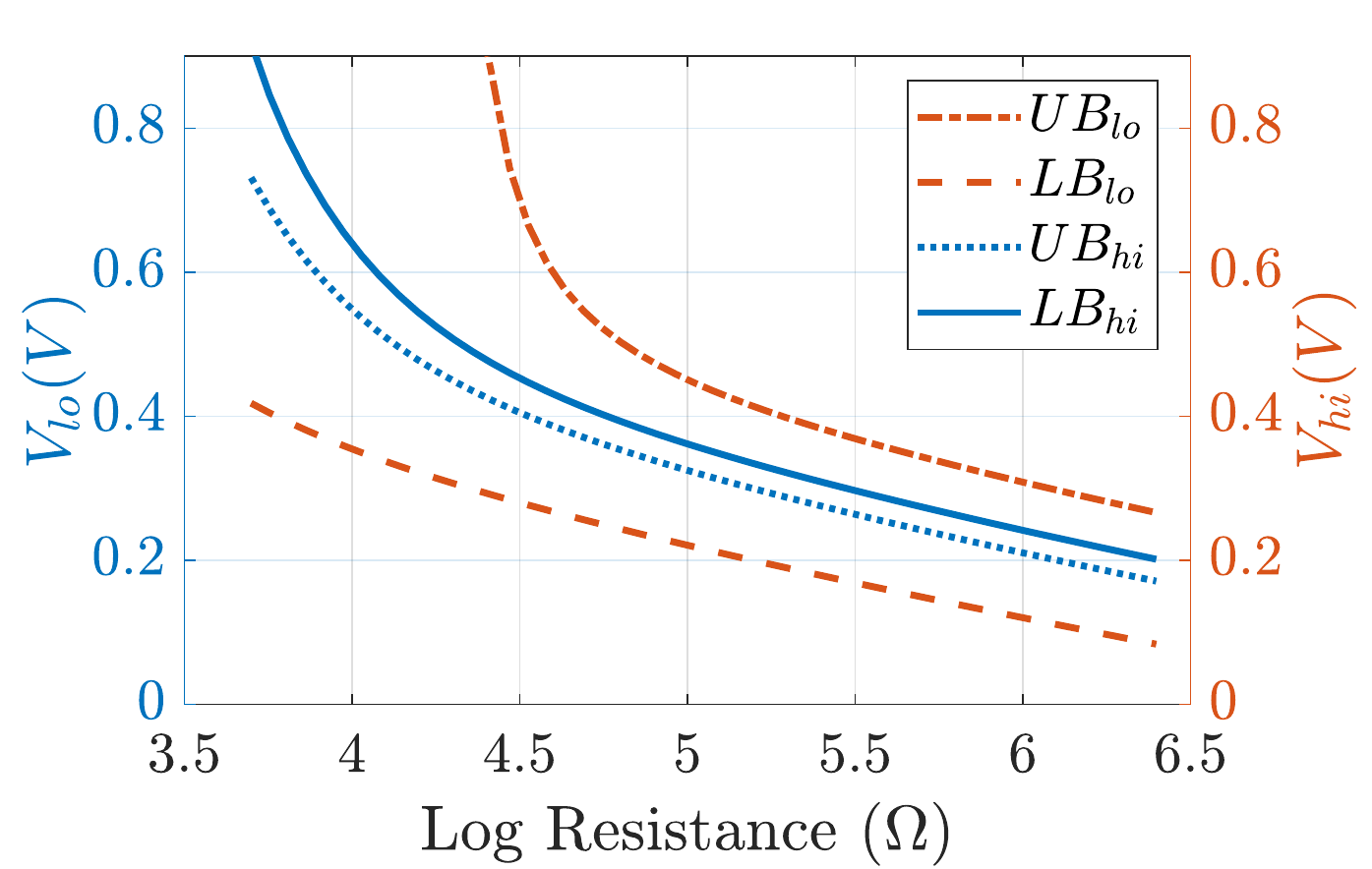}
    \caption{The different bounds plotted as function of the corresponding memristor values for the 6T2M circuit.}
    \label{fig:BDC}
\vspace{-0.15in}
\end{figure}

\textbf{1) Preparatory Phase:} Its objective is to identify for each memristor value a corresponding bound values. These values will be useful in the next phase to determine the intervals. Hence, we first generate the LBS and UBS VTC, i.e. $V_{G1}(R_{LB}, \ V_{DL})$ and $V_{G2}(R_{UB}, \, V_{DL})$ curves, by sweeping UB and LB subcircuits for different $R$ and $V_{DL}$ values similar to Fig. \ref{fig:VG vs VDL}. We capture from the VTC curves for each $R_{LB}$ and $R_{UB}$ value the corresponding $LB_{lo(hi)}$ and $UB_{lo(hi)}$ values. Fig. \ref{fig:BDC} presents the corresponding plots that demonstrate the trend of the different bounds as function of the memristor values. Hereon, we use two look-up-tables to store the resistance and corresponding bounds as triplets as $LUT_{{LB}} = \{(R_{LB}, \, LB_{lo}, \, LB_{hi})\}$ and $LUT_{{UB}} = \{(R_{UB}, \, UB_{lo}, \, UB_{hi})\}$. These tables are sorted by the resistance values in decreasing order.

\textbf{2) Building the intervals Phase:}
 
\begin{figure} [!t]
\centering
\includegraphics[width=1\columnwidth]{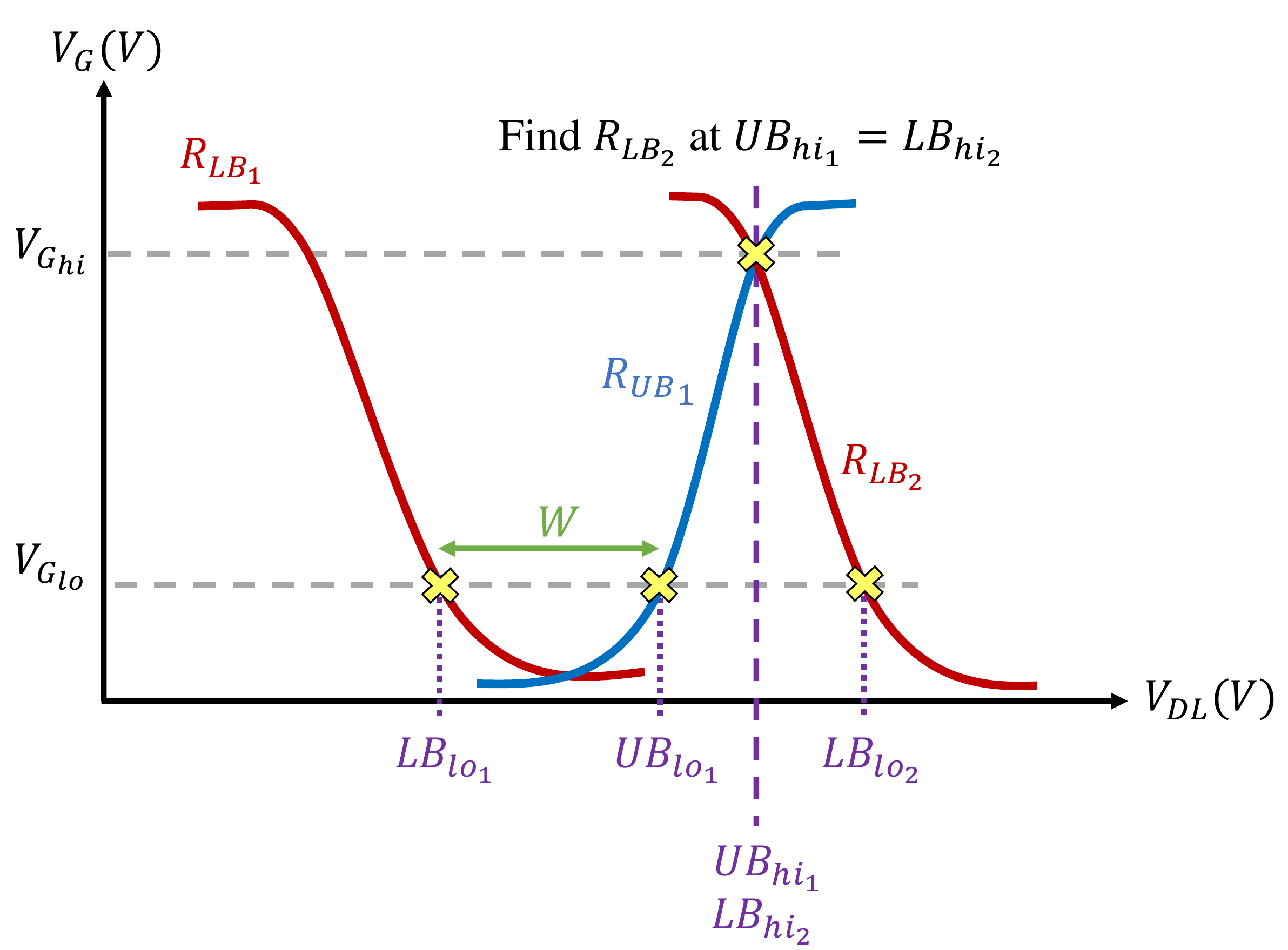}
\caption{Interval building relies on the lookup tables and forbidden regions to identify the next interval.}
\label{fig:interval building}
\vspace{-0.15in}
\end{figure}

Algorithm \ref{tab:interval building pseudocode} presents the pseudo-code for building the intervals. The basics are as follows. The first interval LB is determined based on the lowest possible lower bound value. For a given $LB_{lo}$ value, the corresponding $UB_{lo}$ is determined based on a predefined interval width $W$, and the corresponding $R_{LB}$ and $R_{UB}$ values are determined from the LUTs. To determine the next interval, hereon, while guaranteeing that forbidden regions do not intersect with the chosen intervals, Eq. (\ref{eq1}) must hold. For our purposes, we enforced the conservative requirement presented in Eq. (\ref{eq2}) as illustrated in Fig. \ref{fig:interval building}. As such, we derive $LB_{hi}$ for the next interval from $UB_{hi}$ from the previous one. The corresponding $LB_{lo}$ and $R_{LB}$ can be determined accordingly from the LUTs, and $UB_{(lo, \, hi)}$ and $R_{UB}$ can be determined as stated earlier:

\begin{subequations}
\begin{equation}
\label{eq1}
UB_{lo}^{(i)} \leq UB_{hi}^{(i)}, \, LB_{hi}^{(i+1)} \leq LB_{lo}^{(i+1)}
\end{equation}
\begin{equation}
\label{eq2}
UB_{hi}^{(i)} = LB_{hi}^{(i+1)}
\end{equation}
\end{subequations}

\begin{figure} [!b]
\vspace{-0.1in} 
    \centering
        \includegraphics[width=0.9\columnwidth]{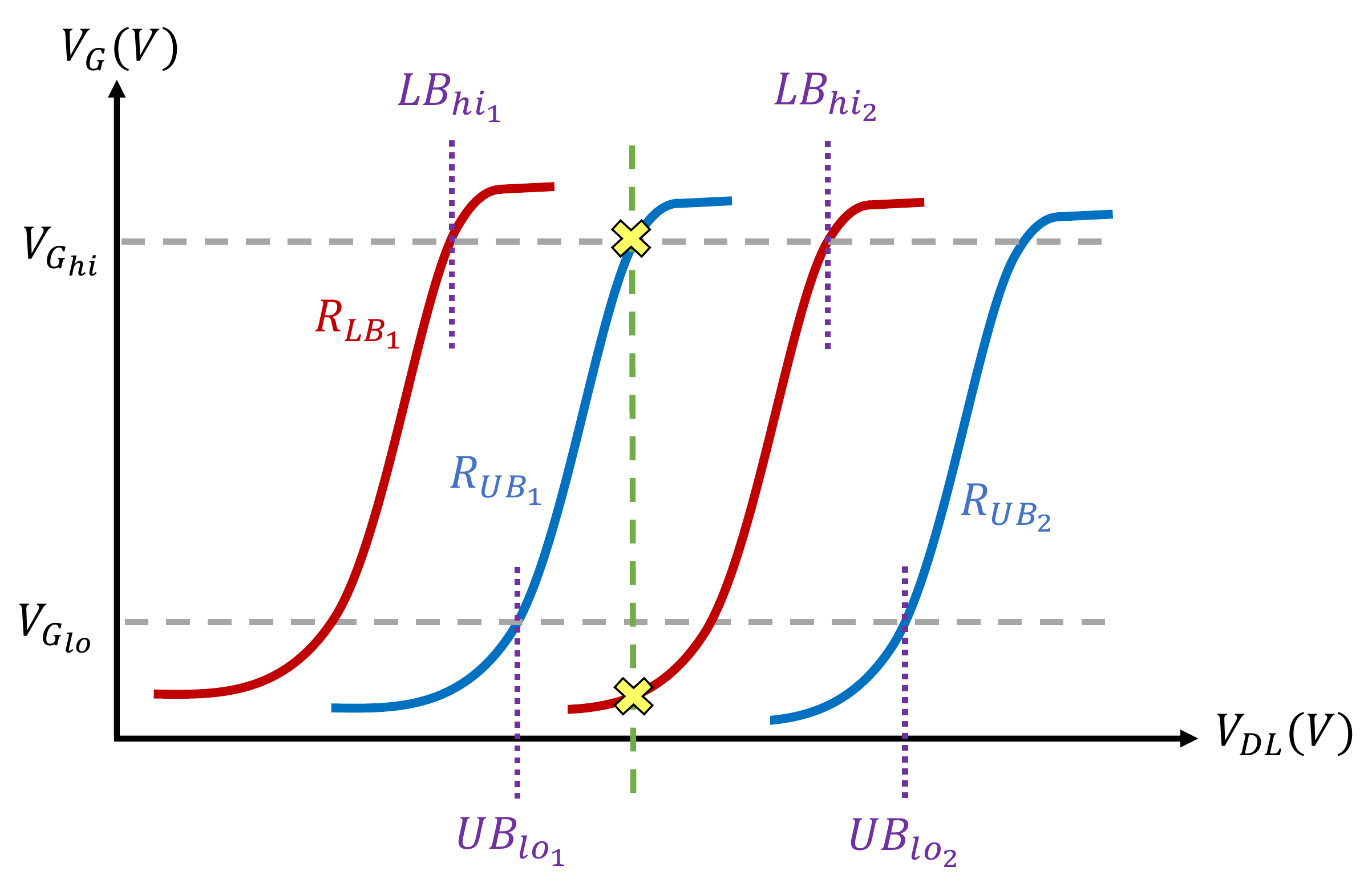}
    \caption{8T2M circuit interval building.}
    \label{fig:8T2M interval}
\end{figure}

The intervals are built in a similar manner for the 8T2M circuit since the UBS does not change. As for the LBS, since the PMOS is an active low device, we exchange $LB_{lo}$ and $LB_{hi}$ so that the matching levels $\{V_{DL}\}$ are now restricted within values $[LB, \, UB]=[LB_{hi}, \, UB_{lo}]$, and the forbidden region becomes: $[LB_{lo}, \, LB_{hi}] \cup [UB_{lo}, \, UB_{hi}]$ as illustrated in Fig. \ref{fig:8T2M interval}. For generic non-symmetrical values for $V_{G_{lo}}$ and $V_{G_{hi}}$, we rely on $[1-\frac{V_{G_{lo}}}{V_{DD}}, \, 1-\frac{V_{G_{hi}}}{V_{DD}}]$.

\begin{algorithm} 
\SetKwInput{KwInput}{Input}                
\SetKwInput{KwOutput}{Output}              
\DontPrintSemicolon
\KwInput{$LUT_{LB}, \, LUT_{UB}$, and $W$.}
\KwOutput{$R_{{LB}_i}, \, R_{{UB}_i}$, and ${D_i}$.}

\tcc{Find the lowest possible $LB_{lo}$ value for $V_{DL}$ and its corresponding resistance}

$[R_{LB}(1), \, LB_{lo}(1)] = LUT_{LB} (1, \, 1:2)$                         

\tcc{Find corresponding $UB_{lo}$ based on the desired interval width $W$}

$UB_{lo}(1)=LB_{lo}(1)+W$
  
\tcc{Search for the memristor value $R_{UB}$ that results in this UB}

$R_{UB}(1)$ = $FindRVal$ ($LUT_{UB}, \, UB_{lo}=UB_{lo}(1))$

$i=1$ 

\While{$R_{UB}(i) > min(R_{UB})$}
{

\tcc{For the same $R_{UB}$ find the corresponding $UB_{hi}$ value}

$UB_{hi}(i)$ = $FindUB_{hi}Val$($LUT_{{UB}}, \, R_{UB}=R_{UB}(i))$                     

$i=i+1$

\tcc{For the next interval} 
     
\tcc{Find the $R_{LB}$ of the next interval that satisfies intersecting boundaries of the forbidden regions  }

$R_{LB}(i)$ = FindRVal($LUT_{LB}, \, LB_{hi}=UB_{hi}(i-1)$)

$LB_{lo}(i)$ = Find$LB_{lo}$Val($LUT_{{LB}}, \, R_{LB}=R_{LB}(i))$             

$UB_{lo}(i)=LB_{lo}(i)+W$

$R_{UB}(i)$ = FindRVal($LUT_{{UB}}, \, UB_{lo}=UB_{lo}(i))$

$UB_{hi}(i)$ = Find$UB_{hi}$Val($LUT_{{UB}}, \, R_{UB}=R_{UB}(i))$   

}

\tcc{Calculate $\mathbf{D}$}             

$\mathbf{D}=(\mathbf{LB}_{lo} + \mathbf{UB}_{lo})/2$                   

\caption{Interval building pseudo-code} 
\label{tab:interval building pseudocode} 
\end{algorithm}

\subsection{Figure of Merit}
\begin{figure} [!t]
\vspace{-0.15in}
\centering
\includegraphics[width=0.9\columnwidth,height=0.6\columnwidth]{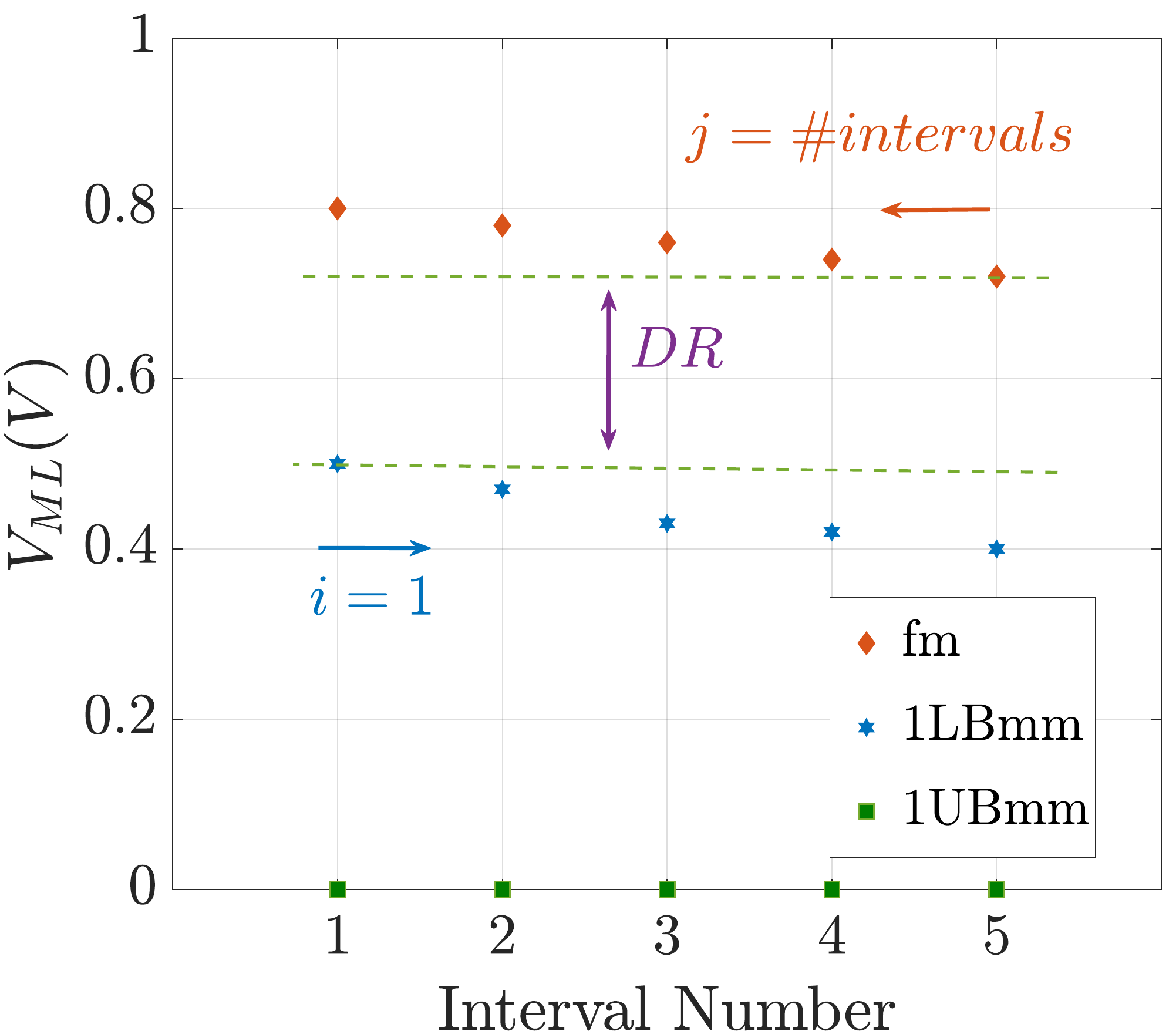}
\caption{$V_{ML}$ for all intervals given a match, one lower bound mismatch, and one upper bound mismatch.}
\label{fig:DR}
\end{figure}

In order to have a fair comparison between the different circuits, we deﬁne a Figure of Merit, $FOM$, which is a function of the following three important metrics:
\begin{itemize}
\item Latency time, $T$, which is the sampling time at which we measure the match line voltage ($V_{ML}$).
\item Number of desired intervals, $\kappa$,  where $2\leq\kappa\leq\eta$ and $\eta$ is defined by the algorithm.
\item Dynamic Range, $DR$, is the maximum possible separation between the full match and 1-mismatch cases for one combination of ($level, \, N, \, T$), as shown in Fig.\ref{fig:DR}, where $level$ refers to the defined ranges of $VG_{{lo}, \, {hi}}$. $N$ represents the number of cells per row. The dynamic range measurements accommodate for the multiple stored input states $\{D_i\}$  corresponding to the multiple intervals, and hence requires measurements of $V_{ML}$ at different combinations of stored matched and mismatched values.  
\end{itemize}

Hence, to find the $DR$ for all possible intervals $\eta$, we proceed as follows. A cell storing value $D_i$ will result in a match when the input is $D_i$ and a mismatch at all other input states $\{D_j\}$, ($j \neq i$) values where $ i, \, j \in [1, \, \eta]$. For each $D_i$, we measure the ML value for the full-match case, $V_{{fm}_i}$. For the 1-mismatch case, we choose the closest $D_j$ values. Hence, we measure ML for two scenarios: a lower bound mismatch case, $V_{{1LBmm}_j}$ where the mismatch cell is storing $D_j$ at $j=i-1$, and an upper bound mismatching cell case, $V_{{1UBmm}_j}$  where the mismatch cell is storing $D_j$ at $j=i+1$. Assuming that a full-match maintains ML high, the dynamic range can be defined as follows: 
\begin{equation}
DR = min(\{V_{fm}\})-max(\{V_{1LBmm}\}\cup\{V_{1UBmm}\})
\end{equation}


The data was monotonic and for a desired number of intervals $\kappa \in [2, \, \eta]$, we relied on a heuristic to find best $DR_\kappa= max(DR)$ for all possible combinations of $\kappa$ intervals. Hence, the Figure of Merit ($FOM$) for a given ($level$, $N$, $\kappa$) and a desired operating range for the sampling time $T$ is deﬁned as follows:
\begin{equation}
FOM(level, \, N, \, \kappa) = \max\limits_T(\frac{DR_\kappa}{T})
\end{equation}

\subsection{Variability Considerations}
\begin{figure} [!t]
    \centering
        \includegraphics[width=0.9\columnwidth]{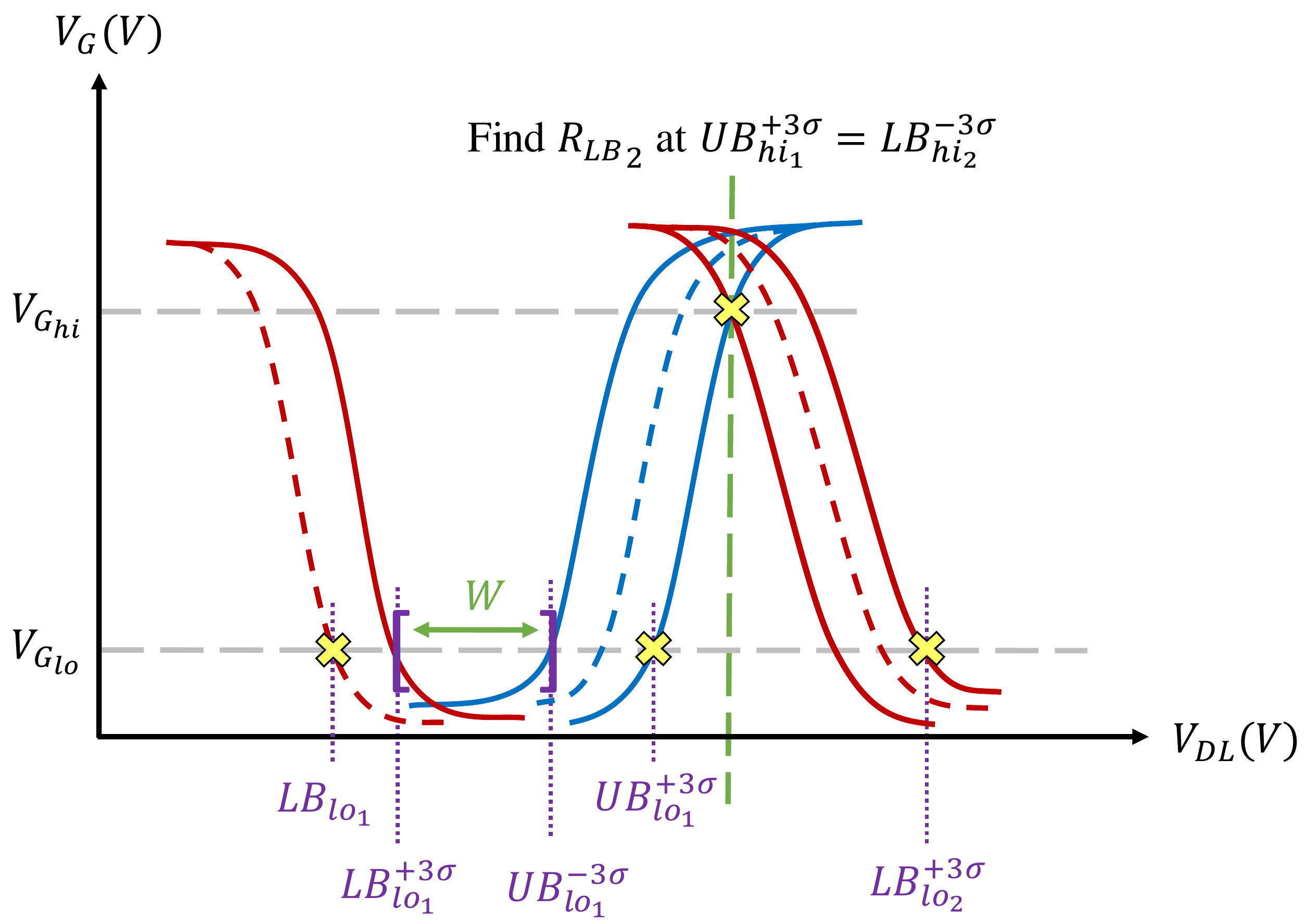}
    \caption{Building intervals with variability.}
    \label{fig:interval building variability}
\vspace{-0.15in} 
\end{figure}

For a given cell with a targeted $R_{LB}$ and $R_{UB}$ pair, the lower bound and upper bound will vary in the presence of process variations, and accordingly, the same cell with the same R will behave differently. Thus, in this section, our objective is to evaluate the impact of process variations on the intervals and the overall performance of the design. For the different cell instances with the same R values to match, we need to compensate for the variability in the interval building by an additional guard band. To do so, we redefined the lower and upper bounds to include variability based margining as shown in Fig. \ref{fig:interval building variability}.

For purposes of illustration, we rely here on the most conservative 3 sigma bounds. We later generalize and test the performance when guarding for less conservative bounds of the general form of $\pm m\sigma$, where the multiplier $m \in [0; \ 3]$. 
In the new margining, each $R_{LB}$ is associated with a range of values for $LB_{lo}$ between  {$LB_{lo}^{+m\sigma}$ and $LB_{hi}^{-m\sigma}$} {where
\begin{equation}
    LB_{lo}^{+m\sigma}=\mu_{LB_{lo}}+ m\sigma_{LB_{lo}}
\end{equation}
and the mean and standard deviation, $\mu_{LB_{lo}}$ and $\sigma_{LB_{lo}}$, of $LB_{lo}$ are obtained from Monte Carlo simulations on the lower bound circuit for the targeted response low voltage level, $V_{G_{lo}}$. The same applies for $LB_{hi}^{-m\sigma}$.} Similarly, each $R_{UB}$ is associated with a value for $UB_{lo}^{-m\sigma}$ and $UB_{hi}^{+m\sigma}$ that are derived {in a similar fashion using Monte Carlo simulations.} 
Thus, according to Table \ref{tab:definitions2}, we get the new interval bounds as: [$LB_{lo}^{+m\sigma}$, \,  $UB_{lo}^{-m\sigma}$]. 
 
The interval building code in Algorithm \ref{tab:interval building pseudocode} will search for the next resistor using the newly derived {curves \{ ($R_{LB}$, $LB_{lo}^{+m\sigma}$), ($R_{LB}$, $LB_{hi}^{-m\sigma}$), ($R_{UB}$, $UB_{lo}^{-m\sigma}$), ($R_{UB}$, $UB_{hi}^{+m\sigma}$) \}, } to obtain the desired intervals.

\section{Results and Analysis}
In this section, we perform a comparative analysis of the 10T2M, 8T2M, and 4T2M2S designs in terms of the $FOM$ analysis, energy, dynamic range, latency, and variability implications.

\begin{figure*} [!t]
    \centering
       \subfigure[]{\includegraphics[width=0.65\columnwidth]{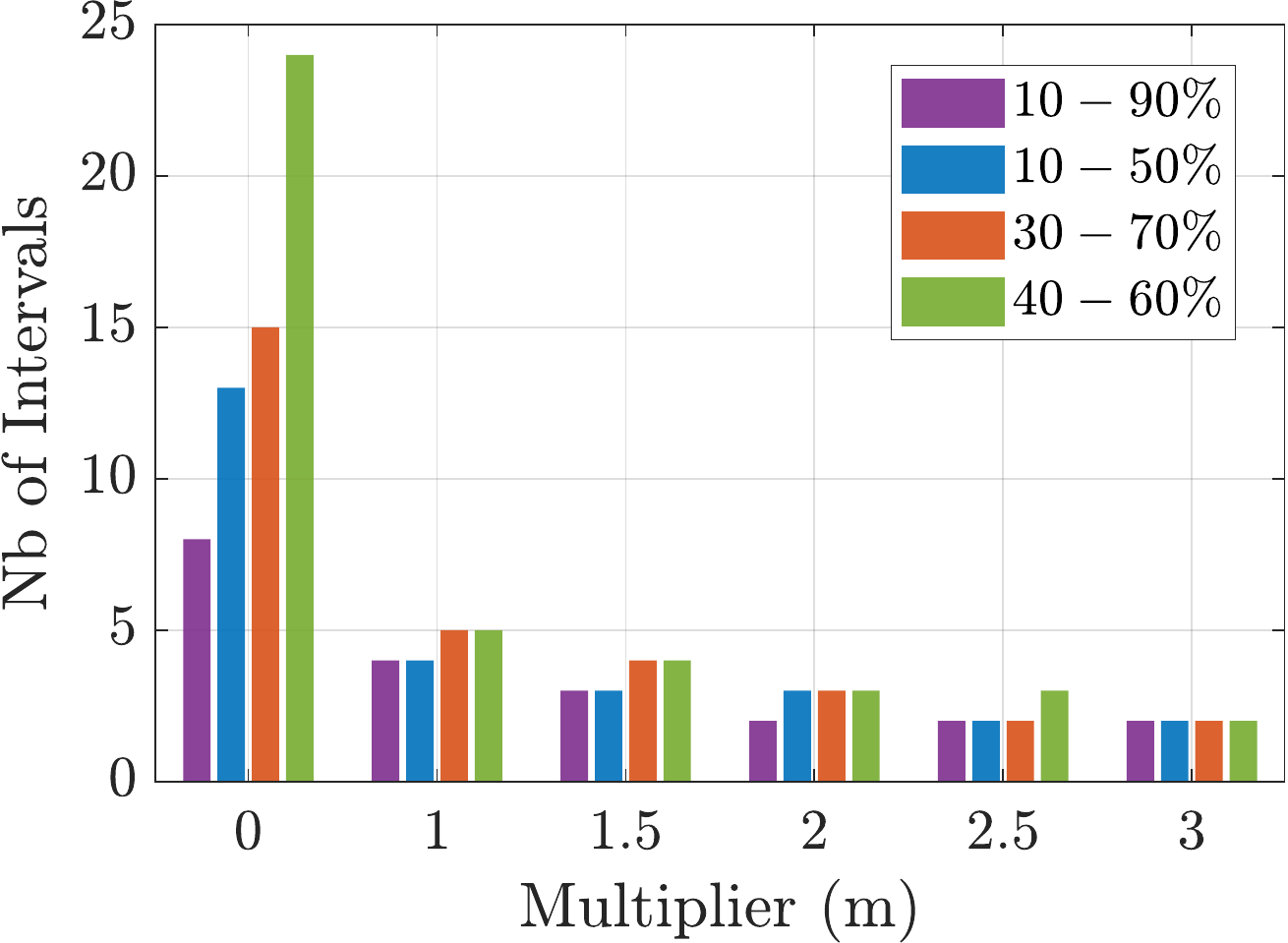}}
       \subfigure[]{\includegraphics[width=0.65\columnwidth]{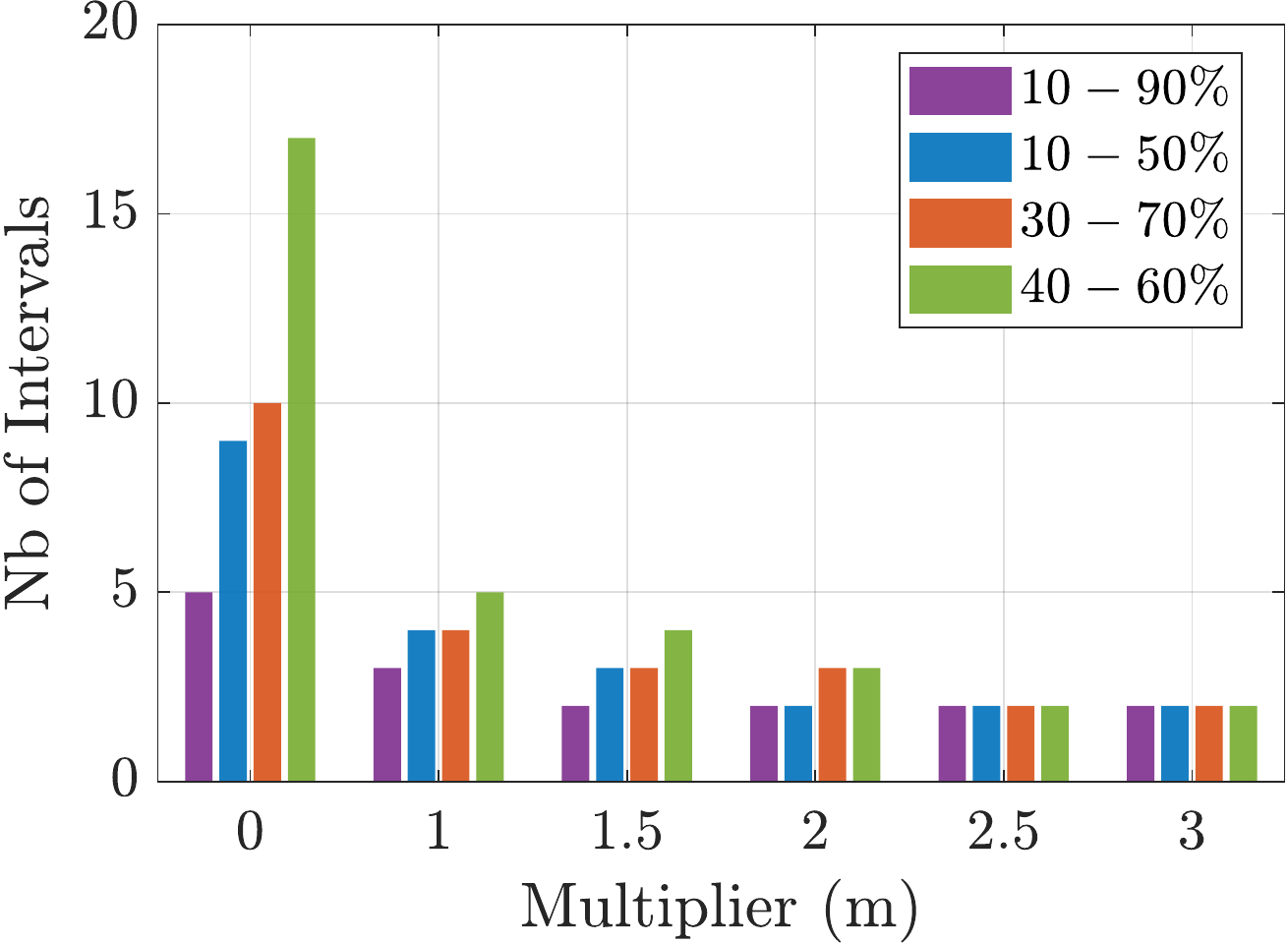}}
       \subfigure[]{\includegraphics[width=0.65\columnwidth]{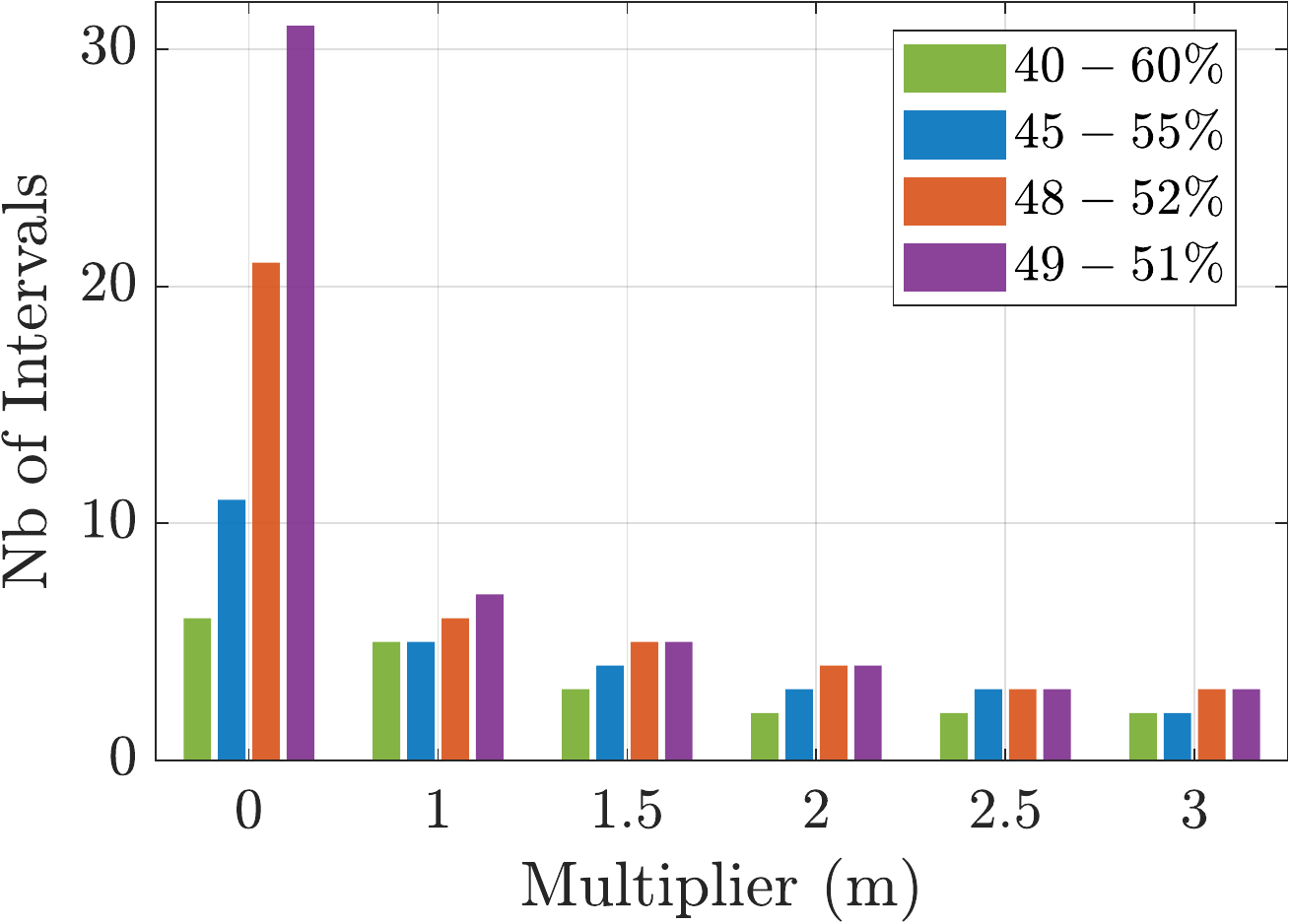}}
    \caption{Number of intervals vs multiplier {(m), as defined in section IV-E}, at different levels for the (a) 10T2M, (b) 8T2M, and (c) 4T2M2S circuits.}
    \label{fig: n vs mult}
\end{figure*}

\begin{figure*} [!t]
    \centering
       \includegraphics[width=0.65\columnwidth]{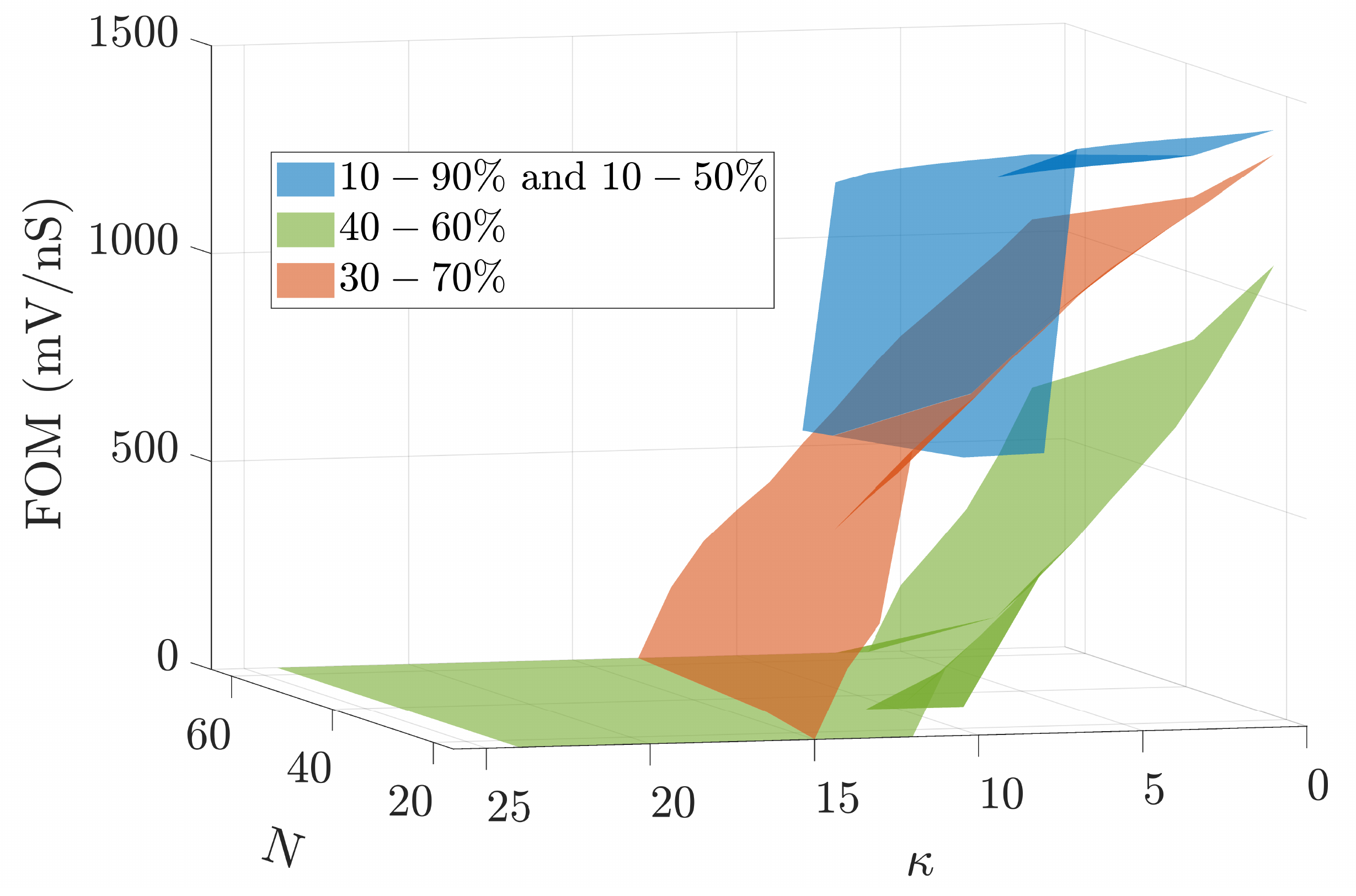}
       \subfigure[]{\includegraphics[width=0.65\columnwidth]{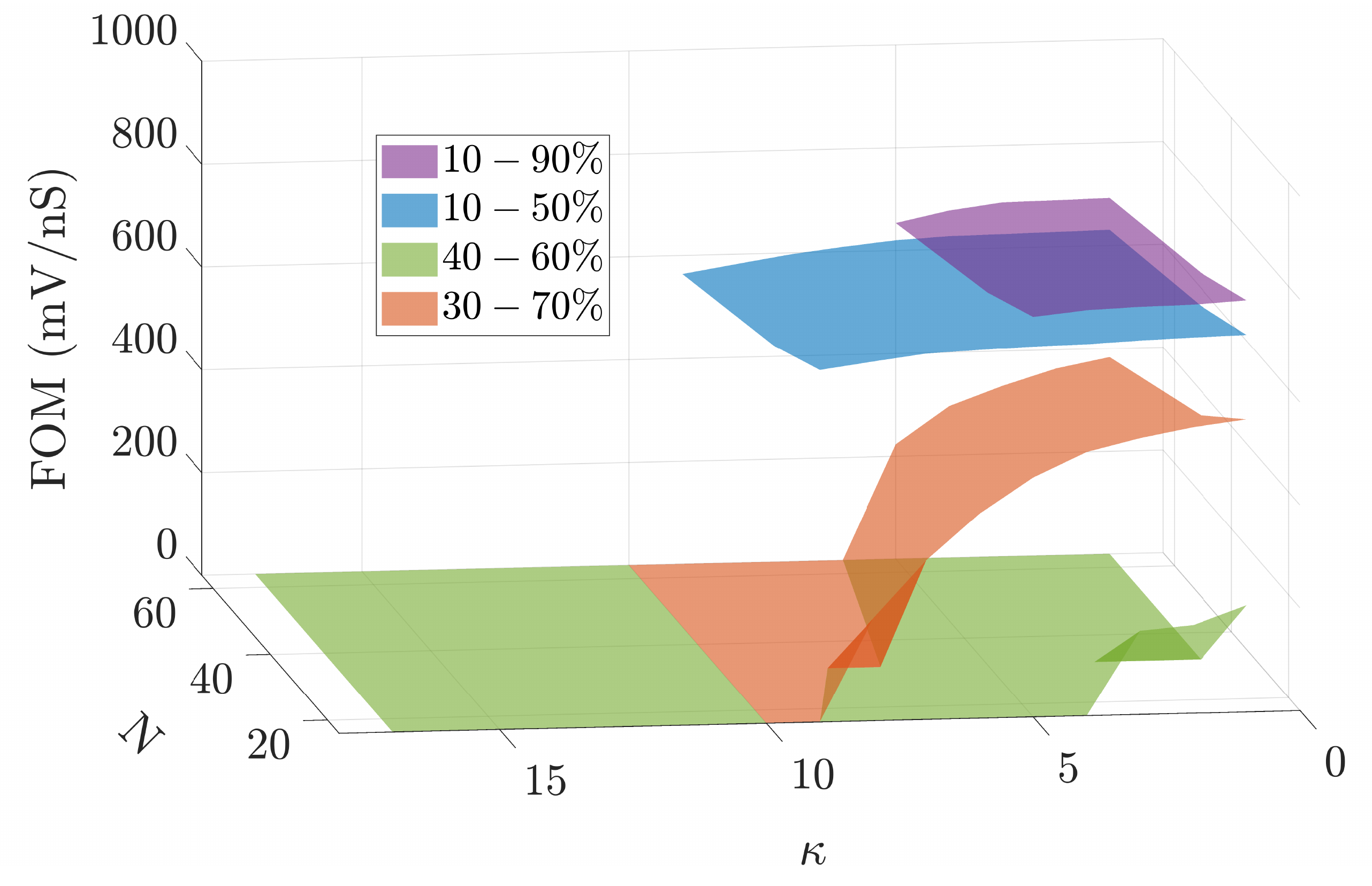}}
       \subfigure[]{\includegraphics[width=0.65\columnwidth]{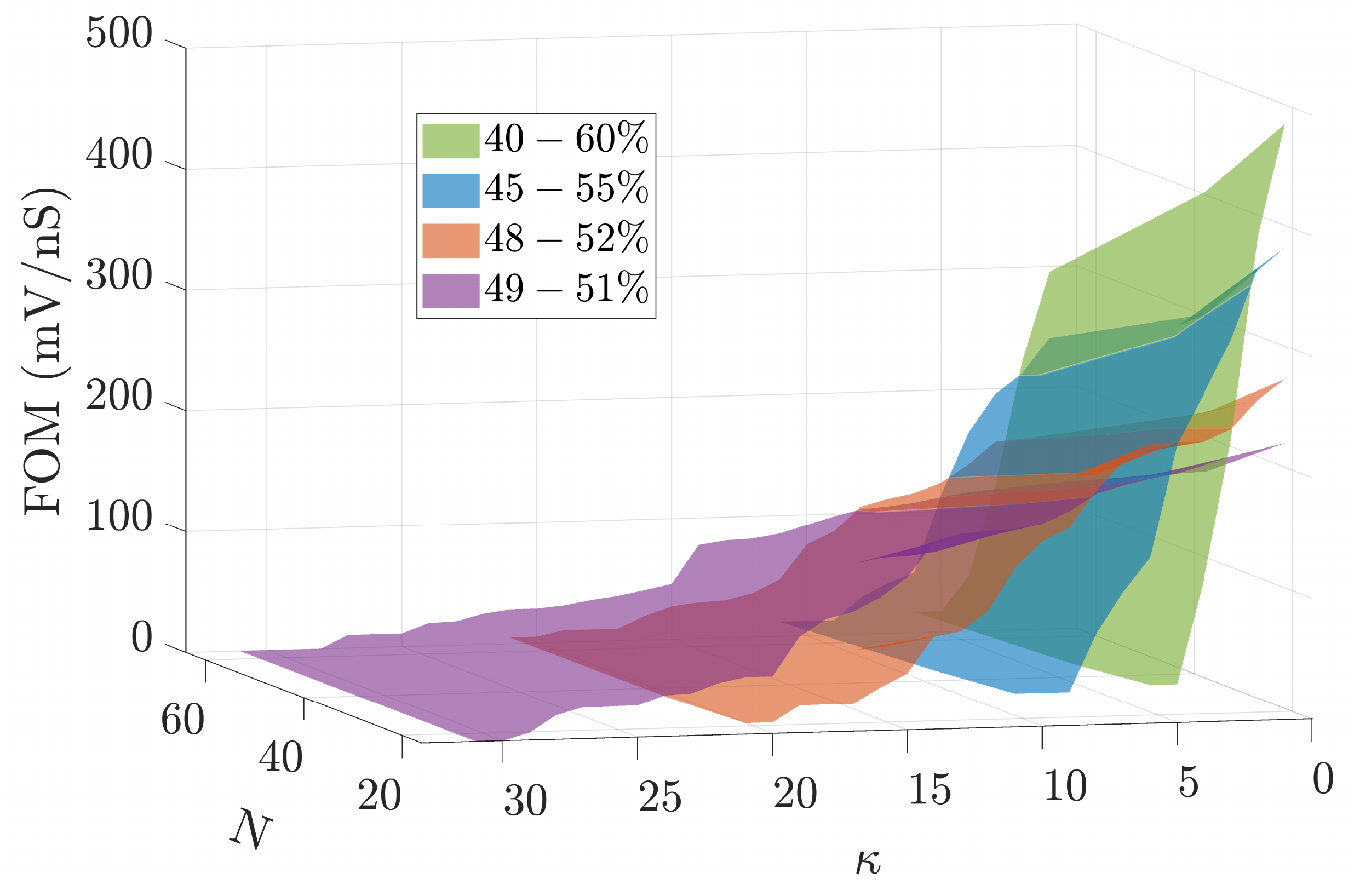}}
    \caption{Effect of changing $N$ and  $\kappa$ on the FOM for different levels of the (a) 10T2M, (b) 8T2M, and (c) 4T2M2S circuits.}
    \label{fig:DR 3D}
    \vspace{-0.15in}
\end{figure*}

\subsection{Experimental Setup}
We performed our simulations in HSPICE. We relied on  predictive technology models for 45nm high-k/metal gate CMOS devices \cite{ptm} to study the design metrics and build the different LUTs. Similar to the assumptions in \cite{main}, we set the transistor threshold voltage $V_{tn}=0.4V.$ We also set $V_{DD}=0.8V.$ We used the threshold switching model described in \cite{TS}. To derive the intervals and dynamic ranges, we implemented our algorithms using MATLAB \cite{MATLAB:R2020b}. 
{For our designs, we assumed Redox based devices \cite{main}; hence, the aCAM comprised of 1T1R branches with a resistance range for $R_{LB/UB} \in [5K\Omega, \, 2.5M\Omega]$. For our simulations, we adopted a linear memristor model, similar to \cite{main, HPtreerev}, because we are only reading and not programming the devices in our experiments. 
} The design space parameters are set as follows. For DC sweeps to develop the LUTs, we set $V_{DL} \in [0.1, \ V_{DD}]$. For building the intervals, we set the interval width $W=10mV$. We also set $N\in\{16,\ 32,\ 64\}$. Finally, we examined the interval building capabilities at different levels of [$V_{G_{lo}}$, $V_{G_{hi}}$] = $V_{DD}$*[$p_{lo}$, \, $p_{hi}$] values. For the 10T2M and 8T2M designs, we chose $p_{lo}-p_{hi} = 10-90\%$ to represent strong operation for match and mismatch, $10-50\%$ to represent strong match and slow mismatch. Intermediate levels such as $30-70\%$ and $40-60\%$ represent weaker match and mismatch but increased number of intervals. For the 4T2M2S, we opted for tighter ranges $40-60\%$, $45-55\%$, $48-52\%$ and $49-51\%$ due to the steep subthreshold swing of the TS switch.

\subsection{FOM}
For different applications, the necessary number of intervals will differ, so we studied the $FOM(level, \, N, \, \kappa)$ for different $\kappa$, $N$, and $level$ values for the different circuits. $T=[0.5-10] ns$ for the 10T2M design, 8T2M circuit, and the 4T2M2S design. Note, that for the 4T2M2S, the DR improved for larger sampling time but this had implications on the figure of merit. In fact, both the 10T2M and the 8T2M provide better DR results at reasonably small sampling times (less than $1ns$), unlike the 4T2M2S design which yields wider DR at higher time. As such, we report for each pair $(N, \, \kappa)$ and $level$ the corresponding $FOM$ values as illustrated in Fig. \ref{fig:DR 3D} for the different cells. 
\\As we can see, there is a trade-off for the three designs between the $FOM$ and $\kappa$. This is attributed to the fact that larger $\kappa$ are subject to more ML costraints as illustrated in Fig. \ref{fig:DR} and hence typically result in lower DRs. The overall pool of intervals to choose amongst is also affected by the level chosen to operate at. For instance, stricter level ranges such as the $10-90\%$ and the $10-50\%$ for the 10T2M and 8T2M designs and the $40-60\%$ and $45-55\%$ for the 4T2M2S design provide higher DR at the cost of less number of intervals. Relaxed levels, however, result in lower DR values but yield more intervals. For example, 31 intervals are attainable for the 4T2M2S design for the level $49-51\%$. Finally, it is worth noting, 4T2M2S uses the low gain LB and UB circuits, and thus it has lower number of intervals at $40-60\%$ compared to the other two designs, however, due to the fact that the TS has high subthreshold swing, we can further push operation to tighter level ranges and attain higher number of intervals.


\subsection{Variability Analysis}
To account for the implications of variability, we apply process variations to the device threshold voltages, such that the NMOS devices are subject to $3\sigma_{vt}$ of $\sim 50mV$ variation. We scale the PMOS device threshold voltage variations according to the device widths. We generated the intervals for $m \in [1; \,3]$ mimicking tight and relaxed margin constraints. For each $m$ value, we performed 1000 Monte Carlo SPICE simulations for the different cells subject to random variations. Fig. \ref{fig: n vs mult} shows the maximum number of achievable intervals $\eta$ that we can get for different design levels for the three circuits as we vary the multiplier $m$. It is clear that as we move towards the more conservative scenarios (more guarding against failure), we obtain fewer intervals. In general, the 4T2M2S design yields the highest number of intervals. In the following section, we introduce a summary table where we comment on the overall performance of the three designs and compare the implications of variability on the probability of fail.

{
\subsection{Corner Analysis}
Aside from random variations, we also explore the impact of the corner analysis on the different designs. We thus studied five corners where the threshold voltages of NMOS and PMOS transistors were skewed by $\pm$10\%. For purposes of our results, TT refers to the typical corners, SF refers to slow NMOS and Fast PMOS and so on. Our simulations show a slight change of $\pm2$ intervals in the number of intervals for a given cell at different corners as illustrated in Table \ref{tab:corners} for $40-60\%$ margins. Other design margins results are reported in the Supplementary Tables I, II, and III, with the FS corner yielding the highest number of intervals and depending on the level the SF offering the lowest number of intervals. 
}

\begin{table}[!b]
\centering
\caption{Number of intervals for $40-60\%$ margin at different corners}
\begin{tabular}{c|c|c|c}
Cell   & 10T2M & 8T2M & 4T2M2S \\ \hline
TT &  24 & 17 & 6  \\ \hline
SS & 24  & 17 &  6  \\ \hline
FF & 23  & 16 &  6  \\ \hline
SF & 22  & 16 &  6  \\ \hline
FS & 25  & 18 &  7  \\ \hline
\end{tabular}
\label{tab:corners}
\end{table}

\begin{table}[]
\caption{Comparison Summary}
\centering
\resizebox{1\linewidth}{!}
{
\begin{tabular}{c|cccccc}
Design & \begin{tabular}[c]{@{}c@{}}DR$^a$\\(mV) \end{tabular}& \begin{tabular}[c]{@{}c@{}}Latency$^b$ \\(ns) \end{tabular}& \begin{tabular}[c]{@{}c@{}}Energy$^c$\\(fJ)\end{tabular}& \begin{tabular}[c]{@{}c@{}} Area$^d$ \\ ($\lambda^2$) \end{tabular}& \begin{tabular}[c]{@{}c@{}}Max Nb of \\ Intervals$^e$  \end{tabular} & \begin{tabular}[c]{@{}c@{}}Fail \\ Probability$^f$ \end{tabular} \\ \hline
10T2M   &   371.2  &  0.083   &   73.3   & 1624  &  24 & 0.007 \\ 
8T2M    &   37.5   &  0.03    &   30.4  & 1400  &  17 & 0.035 \\
4T2M2S  &   209.6  &  0.43    &   913.7   &758   &  6  & 0.0697 \\ \hline
\end{tabular}
}
\label{tab: summary}
\\ {\footnotesize
$^a$Calculated at $1ns$, for three intervals. \\
$^b$Obtained for three intervals and DR=100mV.  \\
$^c$Full-mismatch energy evaluated at $1ns$, for three intervals. \\
$^d$ Active area \\
$^e$Obtained from interval building simulations with no variability. \\
$^f$Calculated at $0.5ns$, for three intervals. \\}
\end{table}

\subsection {Comparative Analysis}
Table \ref{tab: summary} summarizes the results for the three cell designs {using 45nm PTM}. We chose the level $40-60\%$ as the basis of our comparisons since it is common among the three circuits, and we set the number of cells per row to 16. To properly assess the designs, we evaluated each of the following metrics subject to a fixed set of constraints. Thus, for purposes of the dynamic range comparisons, we fixed the time at which we read $V_{ML}$ to $1ns$, and the number of intervals equal to $\kappa=3$, and we reported the best dynamic range for each design.
For the latency, we also set $\kappa=3$, and we identified it as the smallest time at which we achieve a DR of $100 mV$. As for the energy dissipation, we fixed the evaluation time to $1ns$ and reported the average energy for the full mismatch case, while choosing the three intervals that give the best DR for each design. 

For the  10T2M and the 8T2M circuits, the energy comprises pre-charge and evaluate energies, with the evaluate energy being mainly driven by  the  voltage-divider circuits of the resistor-based inverters. For  the  4T2M2S,  the  dissipated energy comprises  the evaluate energy that is mainly composed of the voltage dividers’ energy and the energy needed to charge the ML upon mismatch. 
A more detailed energy analysis showing the full match, one lower bound mismatch, and full mismatch energies for each of the three intervals is present in Supplementary Tables IV, V, and VI. We note that the energy consumption rises as the memristance values decrease due to more leakages for the voltage divider circuits. For the same resistance range, the compare energy is almost the same for the different designs. {For the intervals that provided the maximum dynamic range, the 8T2M followed by the 10T2M consumed lower energies compared to the 4T2M2S.} {The same experiments were repeated using 65nm TSMC technology \cite{tsmc}, and similar trends were reported as illustrated in Supplementary Table XIV.}
{As for the area, Table VI presents the estimated active area for the different designs. The 10T2M consumes most area with 4T2M2S requiring additional area due to the presence of the switches. The details of the area estimation are presented in Supplementary Section 5.}

In order to assess impact of variability on the different designs, we study their respective failure probabilities. For each design, we choose the maximum multiplier value that results in 3 intervals ({$m=2.5$ for the 10T2M circuit}, $m=2$ for the 8T2M circuit, and $m=1.5$ for the 4T2M2S circuit). Here, we fixed $T=0.5ns$. {To estimate the failure probability, we performed 1000 Monte Carlo SPICE simulations for a row of 16 aCAM cells subject to threshold voltage variations with $3\sigma_{vt}$ of ~50mV. We  measured VML for fm, 1LBmm and 1UBmm cases for the different intervals for each Monte Carlo run. Then, we identified the best reference voltage value Vref that provides the best separation between the match and mismatch scenarios in the presence of variability. For the 10T2M and 8T2M designs, we counted a simulation with a match value that is below Vref as a match fail, and a simulation with a mismatch value that is above Vref as a mismatch fail. The opposite holds for the 4T2M2S circuit. We defined the failure probability based on the number of match and mismatch fails obtained from the simulations for all the intervals.}
Supplementary Tables VII, VIII, and IX show the fail probabilities at different sampling times for the 10T2M, 8T2M, and 4T2M2S aCAM designs, respectively. {Table \ref{tab: summary} presents the fail probabilities for the three designs at the 40-60\% level and T=0.5ns;} $V_{ref}=100$, $408$ and $123 mV$ for the 10T2M, 8T2M and 4T2M2S designs, respectively.




{As far as the specific metrics studied in Table \ref{tab: summary}, overall, we observe that the 10T2M stands as a viable solution as applicable to the specified metrics combined in terms of BER, DR, latency and energy. The 8T2M maintained the lowest energy consumption, and best latency with a good BER. We note that it also offers higher DR at lower sampling times.  This makes both 8T2M and 10T2M good candidates for AP  and neuromorphic computing applications for which energy metric is favored over latency. Routing applications, where energy is not a concern, latency and BER are critical. 8T2M and 10T2M are good candidates. If latency is not very critical, and we operate with tight switching levels, the 4T2M2S design may be a good candidate as we will discuss next. As such, to further explore the BER, and since accuracy in terms of BER is a key concern for routing applications, we analyzed the probability of failure for the different cells whilst considering all possible switching levels for the different multipliers and number of intervals. Supplementary Table X presents the summary for the combinations that resulted in the lowest $P_f$ values. Our results show that for 3 intervals, the 4T2M2S solution offered the lowest $P_f$ values and these were obtained when operating at 48\%-52\% level. When latency in addition to accuracy is a concern, the 8T2M offers a good solution for such applications.}

{\subsection{Memristor Variability} Throughout our studies we assumed strict programming schemes, and relied on this assumption to perform our studies as a proof of concept into how the interval building algorithms are envisioned to guard against process variations.  We perform in Supplementary Section 8 a study on the impact of memristor process variations, where we assumed 1\%, 5\%, and memristor value dependent distributions that reach upto 30\% for high resistance values based on data presented in Supplementary Reference [4] and \cite{HPtreerev}.  Without loss of generality, we performed a thorough analysis on the impact of these assumptions on the 10T2M cell for all the different levels and $m$ values. Supplementary Table XV presents these results in terms of the number of intervals and $P_f$. We note that with the application of the proposed algorithms, we only noted a slight impact on the results in the presence of this variability.}

\subsection {Impact of Array Parasitics and Signal Timing}
{The latency simulations in Table \ref{tab: summary} did not fully incorporate the array interconnect parasitics. Herein, we study the effect of the array size on the performance of the aCAM in the presence of parasitic effects. As such, we considered the 10T2M, 8T2M and 4T2M2S aCAM array with different numbers of rows and columns, and replicated the latency analysis for the specific sets of three intervals studied in Table \ref{tab: summary}, targeting the cells maintaining a dynamic range of 100mV. 
We evaluated the latency for an NxM array. We incorporated in our analysis the impact of row and column interconnect parasitics, where the row (column) interconnect is divided into M (N) partitions, each having parasitic resistance of $r=1\Omega$ and capacitance $c=1fF$. The number of partitions involved for a specific cell depends on the cell location (row number and column number).
As discussed in Supplementary Note 9, the effect of row parasitics becomes negligible when we set the proper timing requirements for the early arrival of signal $V_{DL}$ as recommended in \cite{main} for the 6T2M cell. As such, we take into consideration the array parasitic effects and accommodate for an early arrival of $V_{DL}$ signal to compensate for the delay of the parasitic network. We set $V_{DL}$ to arrive earlier than $V_{SL_{HI}}$ by $5\tau$, where $\tau$ represents the expected RC time constant of the interconnect network feeding $V_{DL}$ to the inputs of the aCAM cell. Table \ref{tab:array_summary} presents the summary of the latency simulations for the farthest row (N=512 in this example) as function of the number of columns.}\\

\begin{table}[!t]
\centering
\caption{{Latency $(ns)$ for the 10T2M, 8T2M, and 4T2M2S designs under the conditions: $level=40-60\%$, $N=512, \, r=1\Omega, \, c=1fF$, and guarding time=$5\tau$}}

\begin{tabular}{c|ccccc|}
\cline{2-6}
                             & \multicolumn{5}{c|}{Number of columns}                                                                               \\ \hline
\multicolumn{1}{|c|}{Design} & \multicolumn{1}{c|}{1}    & \multicolumn{1}{c|}{16}   & \multicolumn{1}{c|}{32}   & \multicolumn{1}{c|}{64}   & 128  \\ \hline
\multicolumn{1}{|c|}{10T2M}  & \multicolumn{1}{c|}{0.05} & \multicolumn{1}{c|}{0.1}  & \multicolumn{1}{c|}{0.16} & \multicolumn{1}{c|}{0.29} & 0.62 \\ \hline
\multicolumn{1}{|c|}{8T2M}   & \multicolumn{1}{c|}{0.017}     & \multicolumn{1}{c|}{0.042}     & \multicolumn{1}{c|}{0.095}     & \multicolumn{1}{c|}{0.2*}     &   0.13**   \\ \hline
\multicolumn{1}{|c|}{4T2M2S} & \multicolumn{1}{c|}{0.34} & \multicolumn{1}{c|}{1.07} & \multicolumn{1}{c|}{1.91} & \multicolumn{1}{c|}{3.78} & 7.66 \\ \hline
\end{tabular}

\label{tab:array_summary}

{\footnotesize{*Note that because of the LB PMOS slow discharge and leakages due to weak $R_{OFF}$ for $40-60\%$ margins for the full match, the 8T2M cell dynamic range does not exceed *65mV and **40mV when the number of cells/row exceeds 64 and 128 cells respectively.}}
\end{table}

{Furthermore, we studied the impact of parasitics, $r=1\Omega$ and $c=1fF$, on the response of the cell as function of the farthest row in a 4T2M2S aCAM array with two columns. We swept $V_{DL}$ and studied $V_{ML}$ response with respect to the input $V_{DL}$, measured at 4ns. We varied the number of rows $N \in \{8, 16, 32, 64, 128, 256, 512\}$. Without loss of generality, the cells have a match interval $[LB, \, UB] = [0.37, 0.47]V$, corresponding to resistances $[R_{LB}, \, R_{UB}] = [112.7, 20.9]K\Omega$. By setting $V_{DL}$ to arrive earlier than $V_{{SL}_{HI}}$ by $5\tau$, ML evaluates properly for upto 512 rows, as illustrated in Fig. \ref{fig:3D ML,VDL,N}. In addition, Supplementary Figure 7 analyzes the impact of $V_{DL}$ timing with respect to $V_{{SL}_{HI}}$ on the ML response.} 

\begin{figure} []
    \centering
        \includegraphics[width=0.85\columnwidth]{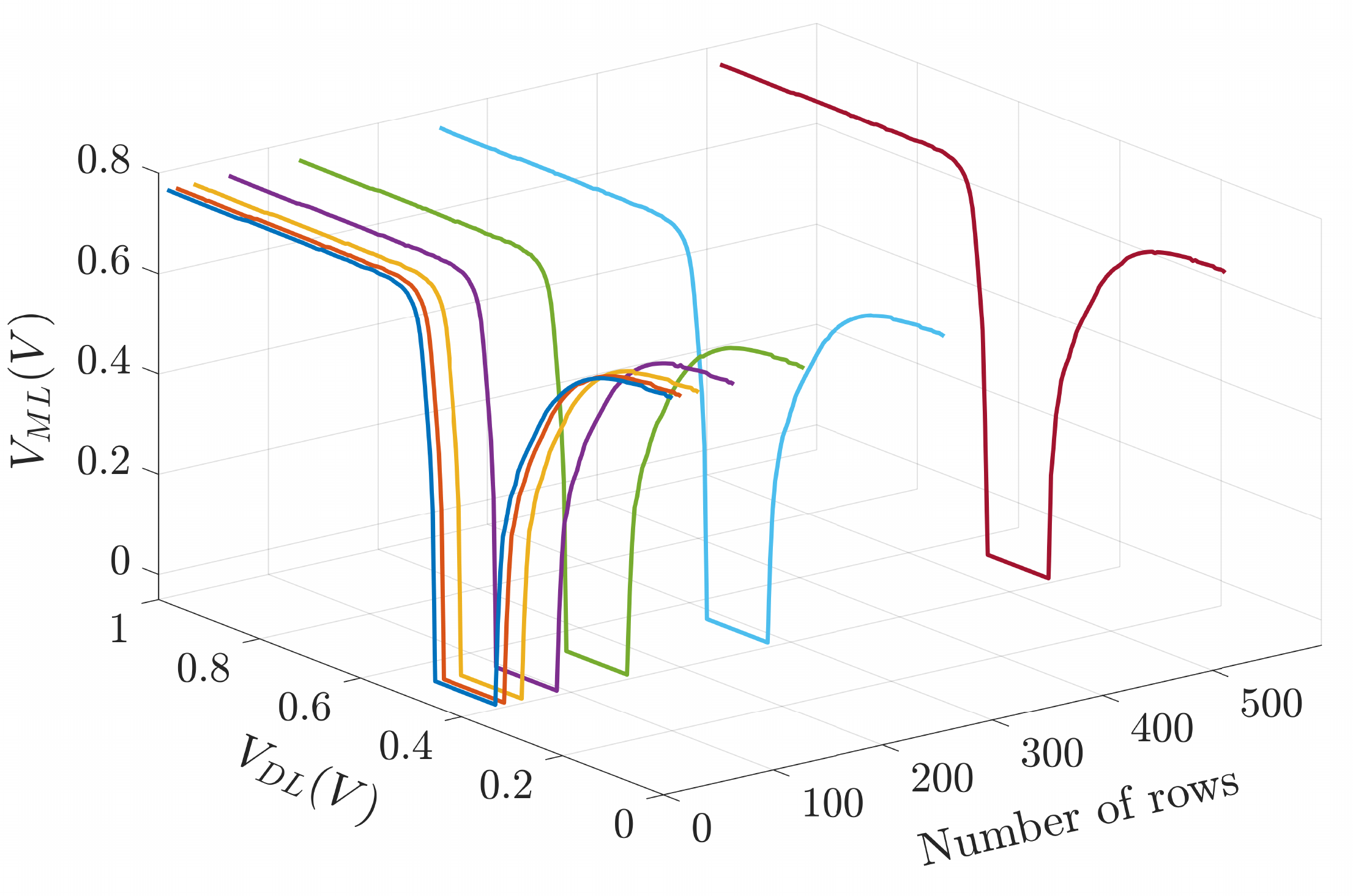}
    \caption{{3D Plot of VML with respect to VDL for an increasing number of rows when VDL arrives earlier than VSLhi by $5\tau$.}}
    \label{fig:3D ML,VDL,N}
\vspace{-0.15in} 
\end{figure}

    
    
    
    
    


\section{Conclusions}
In this paper, we proposed two  new  analog CAM cell designs, the 10T2M and the 8T2M circuits, in an attempt to improve the functionality of the aCAM. We developed the algorithms needed for interval building and Figure of Merit analysis. We performed a comprehensive comparison amongst three different aCAM cell designs: the 10T2M, the 8T2M, and the 4T2M2S aCAM cells and explored the improvements in the gain in comparison to previous work in \cite{main}. We also studied the behavior of the aCAM with variability considerations. The different designs were assessed in terms of key metrics such as the dynamic range, latency, energy, area, maximum number of intervals, and fail probability. The obtained results show a trade-off among the three designs for the different metrics. As such, depending on the desired application, an appropriate aCAM cell design should be adopted.


%





\bibliographystyle{IEEEtran}

\begin{thebibliography}{10}
\bibitem{Kang2012}
S.~M.~S. Kang and S.~Shin, ``Energy-efficient memristive analog and digital
  electronics,'' \emph{Advances in Neuromorphic Memristor Science and
  Applications}, pp. 181--209, 2012.

\bibitem{1083337}
L.~{Chua}, ``Memristor-the missing circuit element,'' \emph{IEEE Transactions
  on Circuit Theory}, vol.~18, no.~5, pp. 507--519, 1971.

\bibitem{8053201}
M.~A. {Bahloul}, M.~E. {Fouda}, R.~{Naous}, M.~A. {Zidan}, A.~M. {Eltawil},
  F.~{Kurdahi}, and K.~N. {Salama}, ``Design and analysis of 2t-2m ternary
  content addressable memories,'' in \emph{2017 IEEE 60th International Midwest
  Symposium on Circuits and Systems (MWSCAS)}, 2017, pp. 1430--1433.

\bibitem{1041061}
H.~J. {Chao}, ``Next generation routers,'' \emph{Proceedings of the IEEE},
  vol.~90, no.~9, pp. 1518--1558, 2002.

\bibitem{253403}
A.~J. {McAuley} and P.~{Francis}, ``Fast routing table lookup using cams,'' in
  \emph{IEEE INFOCOM '93 The Conference on Computer Communications,
  Proceedings}, 1993, pp. 1382--1391 vol.3.

\bibitem{7468531}
C.~{Xu}, S.~{Chen}, J.~{Su}, S.~M. {Yiu}, and L.~C.~K. {Hui}, ``A survey on
  regular expression matching for deep packet inspection: Applications,
  algorithms, and hardware platforms,'' \emph{IEEE Communications Surveys
  Tutorials}, vol.~18, no.~4, pp. 2991--3029, 2016.

\bibitem{patent}
A.~V. {Bechtolsheim} and D.~R. {Cheriton}, ``Access control list processing in
  hardware,'' US Patent 6,377,577, 2002.

\bibitem{main}
C.~Li, C.~E. Graves, X.~Sheng, D.~Miller, M.~Foltin, G.~Pedretti, and J.~P.
  Strachan, ``Analog content-addressable memories with memristors,''
  \emph{Nature communications}, vol.~11, no.~1, pp. 1--8, 2020.

\bibitem{batcher}
K.~E. Batcher, ``Staran parallel processor system hardware,'' in
  \emph{Proceedings of the May 6-10, 1974, national computer conference and
  exposition}, 1974, pp. 405--410.

\bibitem{tracy2016towards}
T.~Tracy, Y.~Fu, I.~Roy, E.~Jonas, and P.~Glendenning, ``Towards machine
  learning on the automata processor,'' in \emph{International Conference on
  High Performance Computing}.\hskip 1em plus 0.5em minus 0.4em\relax Springer,
  2016, pp. 200--218.

\bibitem{guo2011}
Q.~Guo, X.~Guo, Y.~Bai, and E.~{\.I}pek, ``A resistive tcam accelerator for
  data-intensive computing,'' in \emph{2011 44th Annual IEEE/ACM International
  Symposium on Microarchitecture (MICRO)}.\hskip 1em plus 0.5em minus
  0.4em\relax IEEE, 2011, pp. 339--350.

\bibitem{guo2013}
Q.~Guo, X.~Guo, R.~Patel, E.~Ipek, and E.~G. Friedman, ``Ac-dimm: associative
  computing with stt-mram,'' in \emph{Proceedings of the 40th Annual
  International Symposium on Computer Architecture}, 2013, pp. 189--200.

\bibitem{9125991}
F.~Wang, Y.~Feng, X.~Zhan, B.~Chen, and J.~Chen, ``Implementation of data
  search in multi-level nand flash memory by complementary storage scheme,''
  \emph{IEEE Electron Device Letters}, vol.~41, no.~8, pp. 1189--1192, 2020.

\bibitem{materials}
Z.~Wang, H.~Wu, G.~W. Burr, C.~S. Hwang, K.~L. Wang, Q.~Xia, and J.~J. Yang,
  ``Resistive switching materials for information processing,'' \emph{Nat Rev
  Mater}, pp. 173--195, 2020. [Online]. Available:
  \url{https://doi.org/10.1038/s41578-019-0159-3}


\bibitem{9372119}
C.~Li, F.~Müller, T.~Ali, R.~Olivo, M.~Imani, S.~Deng, C.~Zhuo, T.~Kämpfe,
  X.~Yin, and K.~Ni, ``A scalable design of multi-bit ferroelectric content
  addressable memory for data-centric computing,'' in \emph{2020 IEEE
  International Electron Devices Meeting (IEDM)}, 2020, pp. 29.3.1--29.3.4.

\bibitem{9286826}
R.~Rajaei, M.~M. Sharifi, A.~Kazemi, M.~Niemier, and X.~S. Hu, ``Compact
  single-phase-search multistate content-addressable memory design using one
  fefet/cell,'' \emph{IEEE Transactions on Electron Devices}, vol.~68, no.~1,
  pp. 109--117, 2021.

\bibitem{9106766}
X.~Yin, C.~Li, Q.~Huang, L.~Zhang, M.~Niemier, X.~S. Hu, C.~Zhuo, and K.~Ni,
  ``Fecam: A universal compact digital and analog content addressable memory
  using ferroelectric,'' \emph{IEEE Transactions on Electron Devices}, vol.~67,
  no.~7, pp. 2785--2792, 2020.

\bibitem{9474025}
A.~Kazemi, M.~M. Sharifi, A.~F. Laguna, F.~Müller, R.~Rajaei, R.~Olivo,
  T.~Kämpfe, M.~Niemier, and X.~S. Hu, ``In-memory nearest neighbor search
  with fefet multi-bit content-addressable memories,'' in \emph{2021 Design,
  Automation Test in Europe Conference Exhibition (DATE)}, 2021, pp.
  1084--1089.

\bibitem{ref3}
R.~Midya, Z.~Wang, J.~Zhang, S.~E. Savel'ev, C.~Li, M.~Rao, M.~H. Jang,
  S.~Joshi, H.~Jiang, P.~Lin \emph{et~al.}, ``Anatomy of ag/hafnia-based
  selectors with 1010 nonlinearity,'' \emph{Advanced Materials}, vol.~29,
  no.~12, p. 1604457, 2017.

\bibitem{ptm}
\url{http://ptm.asu.edu/}.

\bibitem{TS}
J.~Bazzi, M.~E. Fouda, R.~Kanj, and A.~M. Eltawil, ``Threshold switch modeling
  for analog cam design,'' in \emph{2020 32nd International Conference on
  Microelectronics (ICM)}.\hskip 1em plus 0.5em minus 0.4em\relax IEEE, 2020,
  pp. 1--4.

\bibitem{MATLAB:R2020b}
MATLAB, 2020. [Online]. Available: \url{www.mathworks.com/}

\bibitem{HPtreerev}
G.~Pedretti, C.~E. Graves, S.~Serebryakov, R.~Mao, X.~Sheng, M.~Foltin, C.~Li,
  and J.~P. Strachan, ``Response to reviewers of '' tree-based machine learning
  performed in-memory with memristive analog cam.'','' Accessed: 2022-01-18.
  [Online]. Available:
  \url{https://static-content.springer.com/esm/art\%3A10.1038\%2Fs41467-021-25873-0/MediaObjects/41467\_2021\_25873\_MOESM2\_ESM.pdf}

\bibitem{tsmc}
``Tsmc65nm,'' Accessed: 2022-01-18. [Online]. Available:
  \url{https://www.tsmc.com/english/dedicatedFoundry/technology/logic/l\_65nm}

\end{thebibliography}

\end{document}